\documentclass{article}[a4paper]

\usepackage{siunitx}
\usepackage{subfig}
\usepackage{ulem}
\usepackage{graphicx}
\usepackage[T1]{fontenc}
\usepackage[utf8]{inputenc}
\usepackage[main=english,ngerman]{babel}
\usepackage{amsmath, amsthm, amssymb}
\usepackage{multirow}
\usepackage{siunitx}
\usepackage{pdfpages}

\title{Morphology-Dependent Influences on the Performance of Battery Cells with a Hierarchically Structured Positive Electrode}

\author{
\begin{minipage}{\textwidth}	
	Johanna Naumann,*\textsuperscript{[a]} Nicole Bohn,\textsuperscript{[a]} Oleg Birkholz,\textsuperscript{[b]} Matthias Neumann,\textsuperscript{[c]} Marcus Müller,\textsuperscript{[a]} Joachim R. Binder,\textsuperscript{[a]} Marc Kamlah,*\textsuperscript{[a]} 
\end{minipage}
}

\date{}

\begin{document}
\maketitle

\begin{itemize}	    
	
	\item[{[a]}] J. Naumann*, N. Bohn, Dr. M. Müller, Dr. J.R. Binder, Prof. Dr.-Ing. M. Kamlah*\\
	Institute for Applied Materials, Karlsruhe Institute of Technology, D-76344 Eggenstein-Leopoldshafen, Germany\\
	E-mail: johanna.naumann@kit.edu, marc.kamlah@kit.edu
	
	\item[{[b]}] Dr.-Ing. O. Birkholz\\
	APL Automobil-Prüftechnik Landau GmbH, Am Hölzel 11, D-76829 Landau in der Pfalz, Germany
	
	\item[{[c]}] Dr. M. Neumann\\
	Institute of Stochastics, Ulm University, D-89069 Ulm, Germany
\end{itemize}

\section{Abstract}
The rising demand for high-performing batteries requires new technological concepts. To facilitate fast charge and discharge, hierarchically structured electrodes offer short diffusion paths in the active material. However, there are still gaps in understanding the influences on the cell performance of such electrodes. Here, we employed a cell model to demonstrate that the morphology of the hierarchically structured electrode determines which electrochemical processes dictate the cell performance. The potentially limiting processes include electronic conductivity within the porous secondary particles, solid diffusion within the primary particles, and ionic transport in the electrolyte surrounding the secondary particles. Our insights enable a goal-oriented tailoring of hierarchically structured electrodes for high-power applications.

\section{Introduction}
\label{introduction}

\begin{figure*}[htb]
\begin{center}
\includegraphics[width=\textwidth]{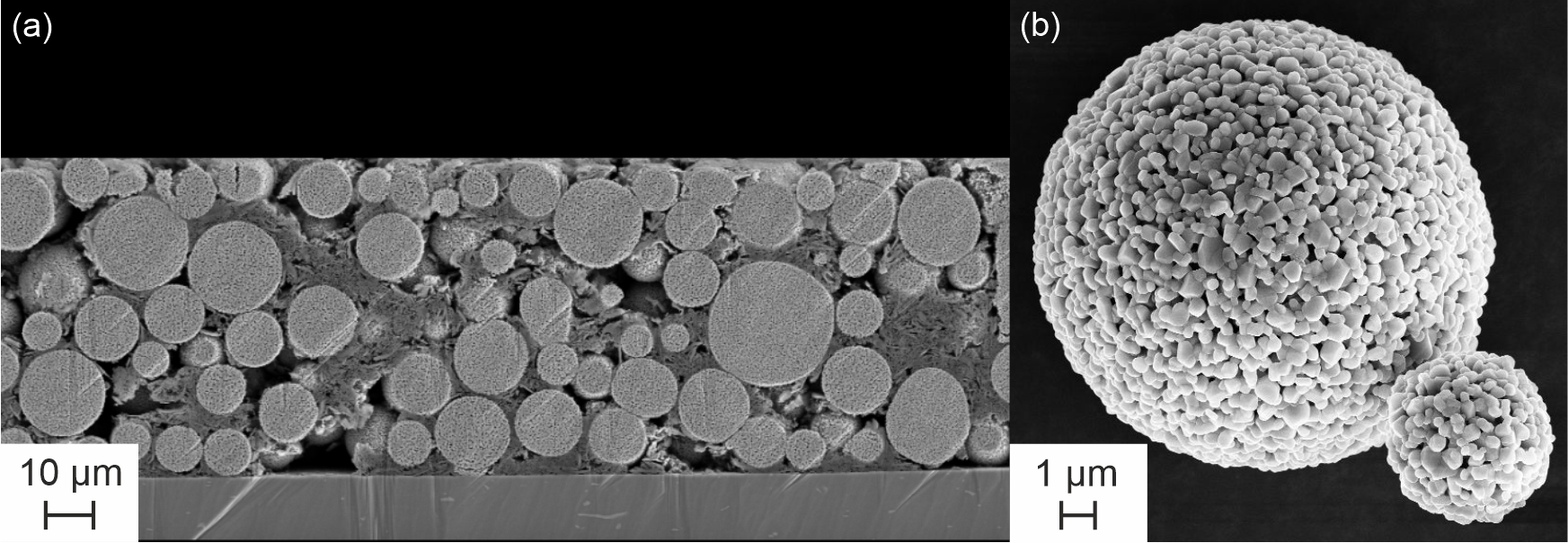}
\caption{SEM images: (a) Cross-section of hierarchically structured electrode E2 with 73~\si{\micro\meter} thickness. (b) Secondary particle of comparable powder to F900 from Wagner et al. \cite{wagner_hierarchical_2020}.}
\label{img:hierelec}
\end{center}
\end{figure*}

The ongoing transition of the energy and transport sector amplifies the requirements on batteries. Fast charging and acceleration of electric vehicles depends on the rate capability of battery cells. One approach to improve the rate capability is the hierarchical structuring of the positive electrode as depicted in \ref{img:hierelec}a. In this case, the secondary active material particles consist of smaller primary particles, where the pores within the secondary particles contain electrolyte, see \ref{img:hierelec}b. Compared to a conventional electrode setup, the electrochemically active surface area of the electrode enlarges and transport paths in the active material shorten. The latter has shown to increase the specific capacity especially at high insertion and extraction
rates \cite{chen_hierarchical_2016, mueller_effect_2021}. The performance of the cell depends on the morphology of the hierarchically structured electrode \cite{wagner_hierarchical_2020}. Despite first experimental studies, the structure-property relationships in hierarchically structured electrodes are not fully understood. Although incorporating the same electrochemical processes as conventional electrode structures, the hierarchical structure affects the interrelation of these processes and additionally requires electronic and ionic transport within the secondary particle. During manufacturing, setting different morphological properties is often interrelated, so separating their influence by individual adjustment remains difficult. Furthermore, experimentally extracting the interplay of morphology and material properties takes great effort. It requires the investigation of a wide range of materials and statistically correlating their inherent characteristics.

To complement experimental investigations, porous secondary particles for battery electrodes have been modeled by several groups. Combining experiments with 3D imaging and statistical image analysis, Wagner et al. \cite{wagner_hierarchical_2020} studied structure-property relationships of hierarchically structured electrodes. Moreover, Neumann et al. \cite{neumann_data-driven_2023} used stochastic nanostructure modeling and numerical simulations to quantitatively investigate effective ionic and electronic transport in secondary active material particles with a designed inner porosity. Wu et al. \cite{wu_mechanical-electrochemical_2016} employed a continuum model of a single secondary particle concentrating on intercalation-induced stresses. Cernak et al. \cite{cernak_three-dimensional_2021} spatially resolved the secondary particle to investigate the effect of secondary particle porosity on the cell performance. Lueth et al. \cite{lueth_agglomerate_2015} developed a continuum cell model for electrodes which consist of secondary particles possessing a certain inner porosity due to their agglomerate character. With this model, they studied the dependence of cell performance on two morphological and two material properties. Namely, they considered effects of the electrochemically active surface area, the secondary particle porosity, the diffusivity in the active material, and the electronic conductivity of the active material. Birkholz et al. \cite{birkholz_electrochemical_2021} applied a similar model to hierarchically structured electrodes with designed inner porosity of the secondary particles. They quantified the influence of diffusivity and electronic conductivity on the rate capability of the cell. However, systematic model-based investigations of the relationship between electrode morphology, material properties, and cell performance across all scales in hierarchically structured electrodes have not been performed so far.

Here, we combined a continuum cell modeling framework with experimental investigations to show that different processes in the battery cell influence its performance depending on the morphology and material properties of the hierarchically structured positive electrode. We determined the change in rate capability in terms of specific capacity with a variety of morphological features as well as their interplay with material properties. Under different conditions, different physical processes within the cell limit its rate capability. From these findings, we deduced recommendations for the design of hierarchically structured electrodes. Our investigations extend the work of Birkholz et al. \cite{birkholz_electrochemical_2021} including a refined parameter set, see Supporting Information section~2. To the best of our knowledge, we are the first group to correlate the influence of morphology and material properties in hierarchically structured electrodes within a cell model. Our contribution provides a basis for a more intentional development of hierarchically structured electrodes.

\section{Experimental}
\label{experimental}

\subsection{Experimental details}

\begin{figure}[hbtp]
	\centering
	\includegraphics[width=\textwidth]{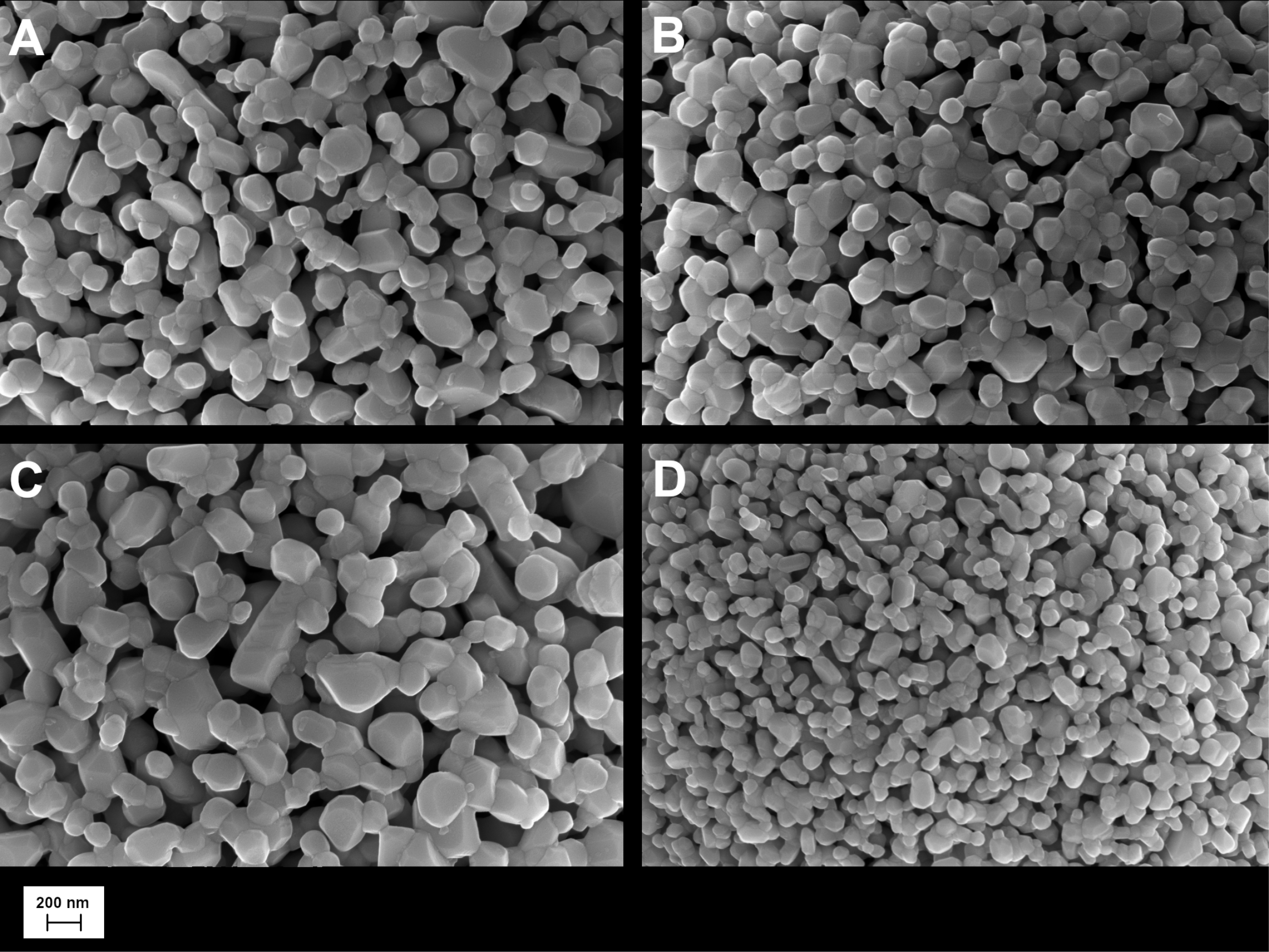}
	\caption{SEM images: NMC111 powders with different secondary particle porosity, electrochemically active surface area and primary particle size. Powders A and C were ground with 0.2~\si{\milli\meter} and powders B and D with 0.1~\si{\milli\meter} grinding balls. The sintering conditions of powders A and B were 800~\si{\degreeCelsius} for 5~\si{\hour}, powder C was sintered at 850~\si{\degreeCelsius} for 5~\si{\hour}, and a two-step sintering process (800~\si{\degreeCelsius} for 1~\si{\hour} and 700~\si{\degreeCelsius} for 10~\si{\hour}) was applied to powder D.\label{img:powders}}
\end{figure}

To compare the modeled data with experimental data, we prepared powders with different secondary particle porosity, electrochemically active surface area, and primary particle size. For this, we milled a commercially available LiNi\textsubscript{1/3}Mn\textsubscript{1/3}Co\textsubscript{1/3}O\textsubscript{2} (NMC111) powder (NM-3100, Toda America) in an agitator bead mill with different grinding media sizes (0.2~\si{\milli\meter} and 0.1~\si{\milli\meter}) and spray dried it to form spherical secondary particles. Afterwards we sintered the two powders at different temperatures (A and B: 800~\si{\degreeCelsius} for 5~\si{\hour}, C: 850~\si{\degreeCelsius} for 5~\si{\hour} and D: 800\si{\degreeCelsius} for 1~\si{\hour} and 700~\si{\degreeCelsius} for 10~\si{\hour}) under air to adjust the primary particle size and porosity (\ref{img:powders}).

We characterized all powders as follows: Secondary particle size by laser diffraction, electrochemically active surface area by BET-method with N\textsubscript{2} adsorption, secondary particle porosity by mercury intrusion, and primary particle size by FE-SEM investigations in combination with image analysis. The detailed processing route and powder characterization was described in Wagner et al. \cite{wagner_hierarchical_2020} and Dreizler et al. \cite{dreizler_investigation_2018}. The results of the porous NMC111 powders are summarized in \ref{tab:powders}.

\begin{table*}[htb]
	\begin{center}
		\caption{Powder characteristics of the porous NMC111 materials.}\label{tab:powders}
		\begin{tabular}{lcccc}	
			\hline		
			\multirow{3}{*}{Electrode} & \multirow{3}{*}{\begin{tabular}{ccc}
					Secondary\\ Particle\\ Porosity
			\end{tabular}} & \multirow{3}{*}{\begin{tabular}{ccc}
					Primary\\ Particles\\ Radius
			\end{tabular}} & \multirow{3}{*}{\begin{tabular}{ccc}
					Electrochemically\\ Active\\ Surface Area
			\end{tabular}} & \multirow{3}{*}{\begin{tabular}{ccc}
					Secondary\\ Particle\\ Radius
			\end{tabular}}\\\\\\
			& \textbf{[-]}	& \textbf{[\si{\micro\meter}]}	& \textbf{[\si{\meter\squared\per\gram}]}	& \textbf{[\si{\micro\meter}]}\\
			\hline
			A 	&	 \textbf{0.514}	&	0.127	&	5.1	&	\textbf{4.5}\\
			B 	&	 \textbf{0.439}	&	0.102	&	4.9	&	\textbf{3.5}\\
			C 	&	 0.476	&	\textbf{0.156}	&	3.6	&	\textbf{4.8}\\
			D 	&	 0.474	&	\textbf{0.072}	&	7.4	&	\textbf{3.6}\\
			\hline	
		\end{tabular}
	\end{center}
\end{table*}

For electrochemical tests, we produced electrodes  by mixing NMC111 powder, PvdF binder (Solef 5130, Solvay Solexis), carbon black (Super C65, Imerys Graphite \& Carbon), and graphite (KS6L Imerys Graphite \& Carbon) in a 87:4:4:5 ratio (wt-\%) in N-methyl-2-pyrrolidone (NMP). We doctor bladed the slurry on an aluminum current collector at 200~\si{\micro\meter} gap height and dried it overnight at 80\si{\degreeCelsius}. After drying we slightly compacted the electrodes. The properties of the electrodes are listed in \ref{tab:elec}. Significant variations in single properties are highlighted in \ref{tab:powders} and \ref{tab:elec}. The following sections describe their effect on the cell performance.

\begin{table*}[htb]
	\begin{center}
		\caption{Properties of hierarchically structured electrodes.}\label{tab:elec}
		\begin{tabular}{lccccc}	
			\hline		
			\multirow{4}{*}{Electrode} & \multirow{4}{*}{Loading} & \multirow{4}{*}{Thickness} & \multirow{4}{*}{\begin{tabular}{cccc}
					Volume Fraction\\ NMC111\\ (w/ internal\\ porosity)
			\end{tabular}} & \multirow{4}{*}{\begin{tabular}{cccc}
					Electrode\\ Porosity\\ (w/o internal\\ porosity)
			\end{tabular}} & \multirow{4}{*}{\begin{tabular}{cccc}
					Volume\\ Fraction\\ Additives
			\end{tabular}}\\\\\\\\
			& \textbf{[\si{\milli\gram\per\centi\meter\squared}]}	& \textbf{[\si{\micro\meter}]}	& \textbf{[-]}	& \textbf{[-]}	& \textbf{[-]}\\
			\hline
			A 						& 7.9 & 48 & 0.3470 (0.7141) & 0.5253 (0.1582) & 0.1277 \\
			B 						& 7.4 & 45 & 0.3609 (0.6433) & 0.5064 (0.2240) & 0.1327 \\
			C 						& 7.4 & 43 & 0.3548 (0.6771) & 0.5147 (0.1924) & 0.1305 \\
			D 						& 8.0 & 50 & 0.3354 (0.6377) & 0.5412 (0.2389) & 0.1234 \\
			E1\textsuperscript{[a]} 	& 8.2 & \textbf{46} & 0.3734 (0.5762) & 0.4893 (0.2865) & 0.1373 \\
			E2\textsuperscript{[b]} 	& 12.4 & \textbf{73} & 0.3559 (0.5991) & 0.5132 (0.2700) & 0.1309 \\
			E3\textsuperscript{[b]} 	& 18.6 & \textbf{107} & 0.3650 (0.6145) & 0.5007 (0.2512) & 0.1343 \\
			E4\textsuperscript{[b]} 	& 22.3 & \textbf{154} & 0.3032 (0.5104) & 0.5853 (0.3781) & 0.1115 \\
			\hline	
		\end{tabular}
	\end{center}
	\footnotesize{\textsf{[a] Comparable powder to F900 from Wagner et al. \cite{wagner_hierarchical_2020} was used. [b] Powder from Dreizler et al. \cite{dreizler_investigation_2018} was used.}}
	
\end{table*}

From the prepared electrodes, a Whatman GF/C separator, electrolyte (LP30), and a lithium metal negative electrode we built Swagelok-type and CR2032 coin-type cells.  For electrochemical testing we used a BT2000 battery cycler from Arbin Instruments. We performed galvanostatic cycling at discharge rates of C/20, C/2, 1C, 2C, 3C, 5C, 7C, and 10C in a voltage range between 4.3 - 3.0~\si{\volt}.

\subsection{Hierarchically structured half-cell model}

For our modeling studies, we used the hierarchically structured half-cell model developed by Birkholz et al. \cite{birkholz_electrochemical_2021}, see Supporting Information section~1. It builds on the electrochemical model for lithium-ion batteries (classical cell model) by Newman and coworkers \cite{doyle_modeling_1993, fuller_simulation_1994, doyle_design_1995, doyle_modeling_1995, doyle_use_1995, doyle_comparison_1996, doyle_analysis_1997, doyle_computer_2000, doyle_computer_2003}. Both models employ the porous electrode theory \cite{newman_porous-electrode_1975, thomas_mathematical_2002, newman_electrochemical_2012}. In this framework, potentials and concentrations within the half-cell are volume-averaged. Only the concentration in the active material is spatially resolved.  The half-cell and the active material particles each represent a pseudo-dimension in the classical cell model. Birkholz et al. \cite{birkholz_electrochemical_2021} extended the classical half-cell model to describe hierarchically structured electrodes (hierarchically structured half-cell model). For this purpose, they defined three pseudo-dimensions: the half-cell level, the secondary particles, and the primary particles. The hierarchically structured half-cell model incorporates the following electrochemical processes:
\begin{itemize}
	\item	Ionic mass and charge transport via the electrolyte both at half-cell level and within the secondary particles.
	\item	Electronic transport in the solid phase both at half-cell level and within the secondary particles.
	\item	Electrochemical reaction at the primary particle surfaces.
	\item	Diffusion of the reduced species in the active material.
\end{itemize}
The solid phase covers the behavior of the active material and the conductive additives. We modified the hierarchically structured half-cell model in two ways. First, we assumed diffusion paths to be longer than the primary particle radius by a factor of 1.5. Schmidt et al. \cite{schmidt_understanding_2021} suggested this assumption if part of the active material particle surface is blocked. This is the case for hierarchically structured electrodes, since the primary particles are sintered together. In the following and the Supporting Information, $R^{(\textrm{I})}$ denotes the representative length of the diffusion paths. Second, we decoupled the electrochemically active surface area of the electrode from other morphological properties by modifying the boundary condition at the primary particle surfaces $r^{(\textrm{I})} = R^{(\textrm{I})}$:

\begin{equation}\label{eq:bc_diffusion}
	\left[D_\text{s}^{(\textrm{I})} \dfrac{\partial c_\text{s}^{(\textrm{I})}}{\partial r^{(\textrm{I})}}\right]_{r^{(\textrm{I})} = R^{(\textrm{I})}} = - \dfrac{\left(R^{(\textrm{I})}/1.5\right) a_\text{s}^{(\textrm{I})}}{3 \varepsilon_\text{s}^{(\textrm{II})}}\bar{j}
\end{equation}

In the original model, the concentration influx $D_\text{s}^{(\textrm{I})} \dfrac{\partial c_\text{s}^{(\textrm{I})}}{\partial r^{(\textrm{I})}}$ equals the species flux density produced by the electrochemical reaction $\bar{j}$. In \ref{eq:bc_diffusion}, we used a prefactor containing the primary particle radius $R^{(\textrm{I})}/1.5$, the volume fraction of the active material within the secondary particles $\varepsilon_\text{s}^{(\textrm{II})}$, and the electrochemically active surface area $a_\text{s}^{(\textrm{I})}$ to enable that they can be chosen independent of each other in a physically consistent manner. Decoupling the electrochemically active surface area from the particle radius within the classical cell model would require a similar approach using the volume fraction of active material in the electrode. We used the extended hierarchically structured half-cell model to study a hierarchically structured positive electrode in a half-cell setup under galvanostatic discharge until a cut-off voltage of $3.0~\si{\volt}$. The corresponding differential equations and boundary conditions were solved in COMSOL Multiphysics.

\subsection{Model validation and morphology variation}

As a reference cell, we chose the half-cell with the E1 electrode. Supporting Information section~2 lists geometric and material properties of the reference cell. Müller et al. \cite{mueller_effect_2021} and Schneider et al. \cite{schneider_transport_2022} investigated similar electrodes, including charge-discharge behavior, incremental capacities (cyclic voltammograms), internal resistances (EIS), and aging behavior.

\begin{figure*}[htb]
	\begin{center}
		\subfloat[]{
			\includegraphics[trim={0 1.3cm 0 0},clip,width=0.5\textwidth]{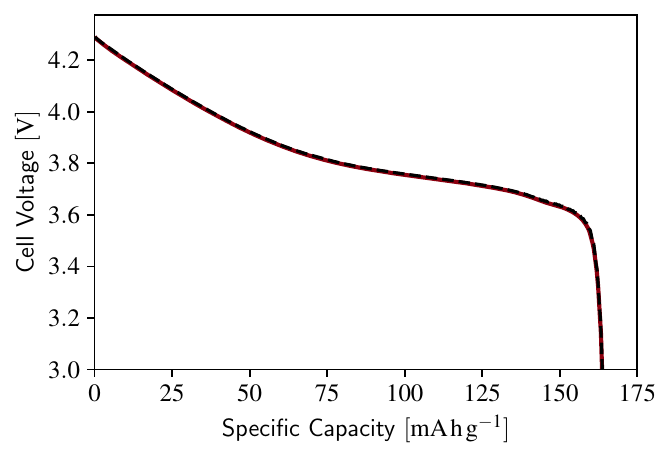}
			\label{img:aprim_C0,05}}
		\subfloat[]{
			\includegraphics[trim={1.5cm 1.3cm 0 0},clip, width=0.435\textwidth]{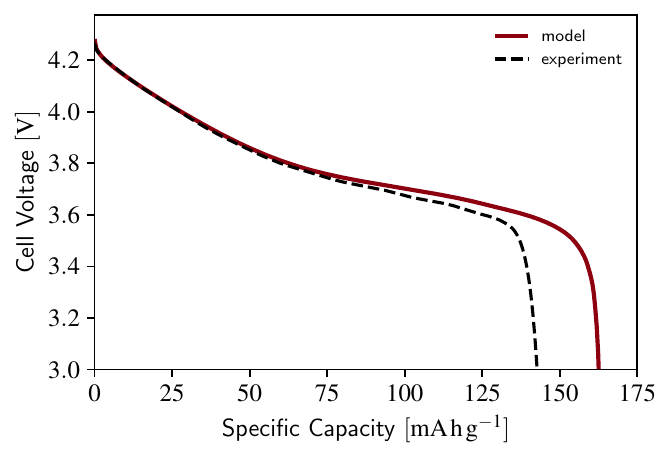}
			\label{img:aprim_C1}}\\
		\subfloat[]{
			\includegraphics[width=0.5\textwidth]{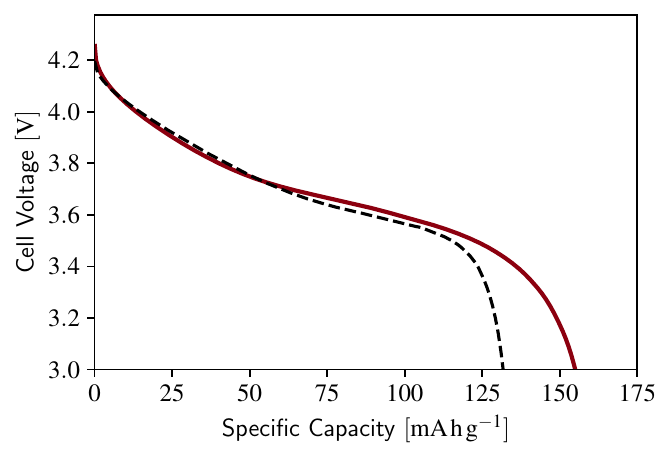}
			\label{img:aprim_C3}}
		\subfloat[]{
			\includegraphics[trim={1.5cm 0 0 0},clip, width=0.435\textwidth]{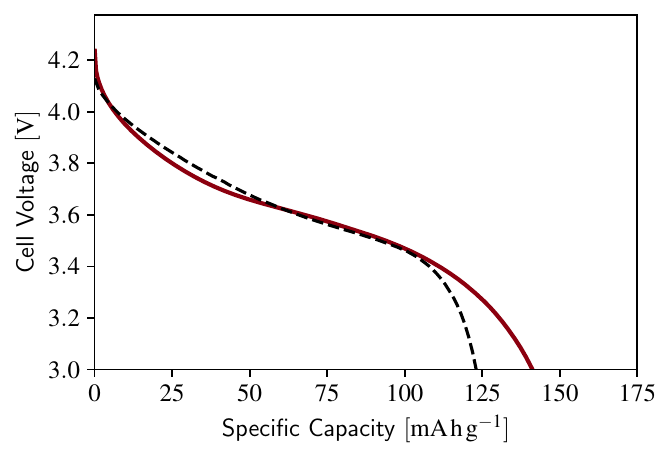}
			\label{img:aprim_C5}}\\
		\caption{Comparison of the modeled and experimental discharge behavior of the reference cell at (a) C/20, (b) 1C, (c) 3C, (d) 5C.}
		\label{img:validation}
	\end{center}
\end{figure*}

To validate the applicability of the hierarchically structured half-cell model, we compared modeled discharge curves at rates between C/20 and 5C to the experimental results. \ref{img:validation} demonstrated their qualitatively good agreement. The model captured the general shape of the discharge curves. In addition, it reproduced the trend of decreasing specific capacity with increasing discharge rate and predicted the specific capacity with an error of $\leq 11.7~\%$. Therefore, we considered the model appropriate to qualitatively study the reference cell and cells with similar electrodes.

With our investigations, we aimed at gaining insight into the structure-property relationships of hierarchically structured electrodes. For this purpose, we separately varied morphological properties and modeled the discharge of the cell at rates ranging from C/20 to 10C. This paper includes studies of the electrochemically active surface area, the secondary particle porosity, the secondary particle size, the primary particle size, and the electrode thickness. For each property, we identified the underlying physical processes which explained the observed behavior of the cell. Besides structural parameters, we modified material properties in order to study their interaction with morphology.

\section{Results and Discussion}
\label{results_discussion}

\subsection{Effect of electrochemically active surface area on reaction kinetics} \label{ch:eas_k0}

\begin{figure*}[htb]
	\begin{center}
		\subfloat[]{
			\includegraphics[width=0.5\textwidth]{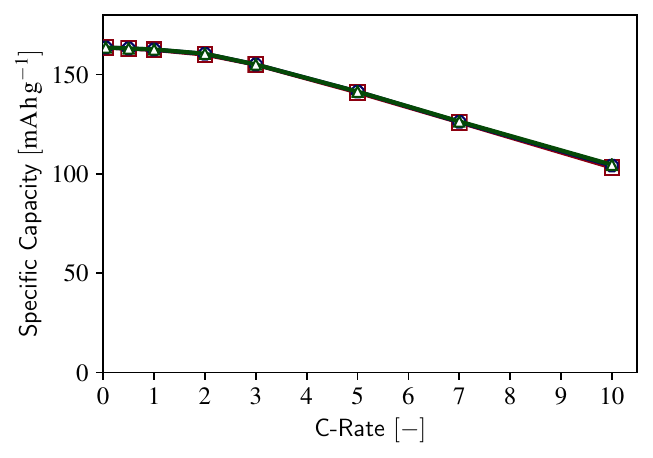}
			\label{img:aprim_C}}
		\subfloat[]{
			\includegraphics[trim={1.55cm 0 0 0},clip, width=0.435\textwidth]{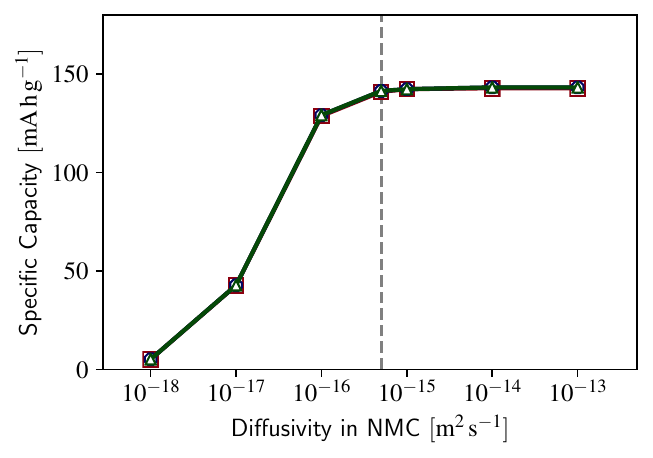}
			\label{img:aprim_Ds}}\\
		\subfloat[]{
			\includegraphics[width=0.5\textwidth]{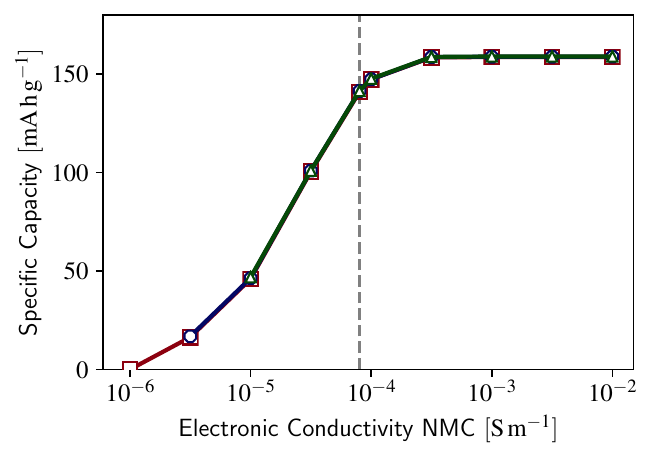}
			\label{img:aprim_sigma}}
		\subfloat[]{
			\includegraphics[trim={1.55cm 0 0 0},clip, width=0.435\textwidth]{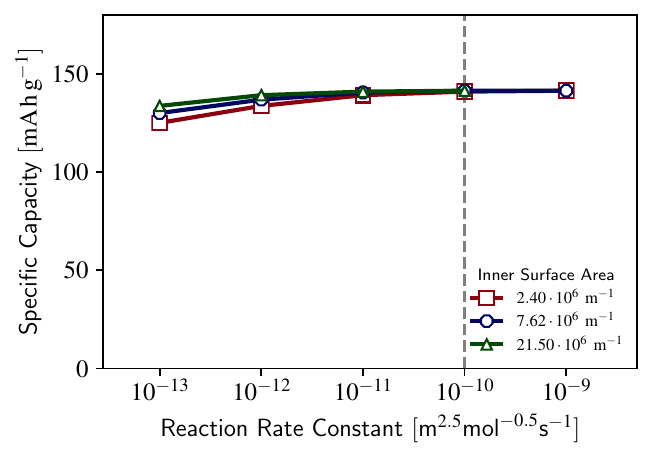}
			\label{img:aprim_k0}}\\
		\caption{(a) Rate capability of electrodes with different electrochemically active surface area. (b-d) Influence of (b) diffusivity in the active material, (c) electronic conductivity of the active material, (d) reaction rate on the cell performance at 5C.}
		\label{img:aprim}
	\end{center}
\end{figure*}

The first goal was to quantify the influence of the electrochemically active surface area on the rate capability. Studying this property experimentally is especially difficult since in reality the electrochemically active surface area depends on primary and secondary particle size as well as on porosity. Our cell model provided a deeper understanding of theses single factors by separating the effect of electrochemically active surface area on the cell performance from the other influencing factors. With an extension of the hierarchically structured half-cell model in \ref{eq:bc_diffusion}, we enabled a variation of the electrochemically active surface area independent from all other electrode properties. In addition to the boundary condition at the primary particle surfaces, the electrochemically active surface area affects the source terms of the hierarchically structured half-cell model (Supporting Information Eq.~4-5 and Eq.~14-16), which describe the magnitude of intercalation. The reference electrode had an electrochemically active surface area of $7.62 \cdot 10^{6}~\si{\per\meter}$. The investigated range of $2.40 \cdot 10^{6}~\si{\per\meter}$ (0.5~\si{\meter\squared\per\gram}) to $21.50 \cdot 10^{6}~\si{\per\meter}$ (4.5~\si{\meter\squared\per\gram}) corresponds to typical values observed in experiments \cite{wagner_hierarchical_2020}. Supporting Information Eq.~33 gives a conversion of \si{\per\meter}, which is required by the hierarchically structured half-cell model, to \si{\meter\squared\per\gram}, which is common in experimental contexts. \ref{img:aprim_C} showed that the acquired rate capability of cells with electrodes of different electrochemically active surface area was identical. Therefore, we find that within the chosen range, the electrochemically active surface area of the electrode has no effect on the cell performance. Lueth et al. \cite{lueth_agglomerate_2015} found the discharge capacity of agglomerates significantly depends on the electrochemically active surface area of the electrode. At the investigated 5C discharge rate, their model yielded a constant decrease in capacity toward the end of discharge. This contrasts with the S-shaped discharge curves from our modeling results, which agreed with the experimental findings, see \ref{img:validation}. The difference in discharge profiles highly affects the obtained capacity and offers an explanation for the deviation between the two models.

To broaden our results, we studied electrodes with different material properties. Namely, we altered material properties which interdepend with the studied morphological properties, i.e. reaction rate, electronic conductivity of the active material, and diffusivity in the active material (see Supporting Information Eq.~6, 16, and 23). \ref{img:aprim_Ds} depicts the dependence of cell performance on the active material diffusivity at 5C as an example for a high discharge rate. For a given diffusivity, electrodes with different electrochemically active surface area yielded identical cell performance. Likewise, varying the electrochemically active surface area did not affect the influence of the electronic conductivity of the active material on the cell performance, see \ref{img:aprim_sigma}. Only reaction rate constants lower than $10^{-10}~\text{m}^{2.5}\text{mol}^{-0.5}\text{s}^{-1}$ caused a difference in rate capability as depicted in \ref{img:aprim_k0}. In this case, high surface area electrodes retained a higher specific capacity at 5C because the rate of the electrochemical reaction determined the cell performance. Since the deviation was small even under these extreme conditions, we conclude that hierarchically structured electrodes in general tend to be insensitive to a change in electrochemically active surface area. 

\subsection{Interplay between secondary particle porosity and electronic conductivity}\label{ch:res_sigma}

\begin{figure*}[htb]
	\begin{center}
		\subfloat[]{
			\includegraphics[width=0.5\textwidth]{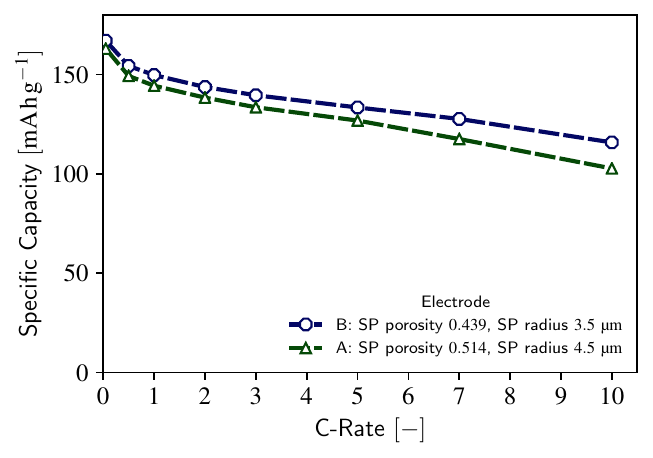}
			\label{img:epssec_exp}}
		\subfloat[]{
			\includegraphics[trim={1.55cm 0 0 0},clip, width=0.435\textwidth]{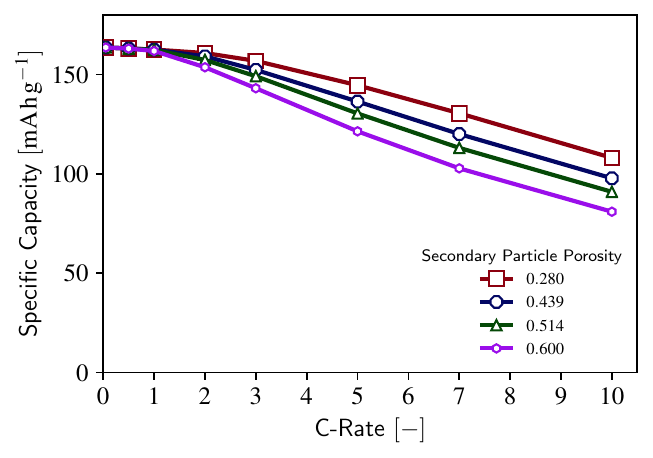}
			\label{img:epssec_sim}}\\
		\caption{(a) Experimental rate capability of electrodes A and B with differing secondary particle porosity and size, see \ref{tab:powders} and \ref{tab:elec}. (b) Modeled rate capability depending on secondary particle porosity.}
		\label{img:epssec}
	\end{center}
\end{figure*}

Next, we investigated to what extent the cell performance of hierarchically structured electrodes depends on the secondary particle porosity. For this purpose, we studied electrodes A and B, which featured differing secondary particle porosity and size but comparable primary particle size. \ref{img:epssec_exp} revealed a decreasing specific capacity with increasing discharge rate. Furthermore, the capacity fade was more pronounced for electrode A containing more porous and larger secondary particles. The hierarchically structured half-cell model could separate the two influences. In this section, we first investigated the influence of secondary particle porosity on the rate capability. For this purpose, we modeled electrodes of varying secondary particle porosity. At the same time, all other electrode properties complied with the reference electrode. The secondary particle porosity enters the hierarchically structured half-cell model in three ways. It explicitly affects the ionic transport (Supporting Information Eq.~14) as well as the effective transport properties (Supporting Information Figure~2) in the secondary particle. Furthermore, it changes the applied current (Supporting Information Eq.~8, 9, and 13) in order to be consistent with C-rate. The hierarchically structured half-cell model reproduced the general discharge and rate behavior as shown in \ref{img:epssec_sim}. Hence, the modeling results could be used to qualitatively explain the processes during discharge. In addition, the model confirmed the experimental trend with respect to secondary particle porosity. More porous particles provided a lower rate capability, which explained part of the deviation between electrodes A and B.

\begin{figure*}[htb]
	\begin{center}
		\subfloat[]{
			\includegraphics[trim={1.2cm 0 0 1.0cm},clip, width=0.5\textwidth]{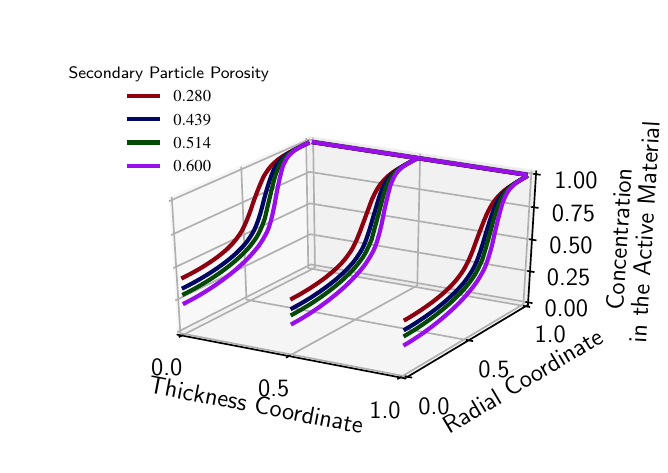}
			\label{img:epssec_cs}}
		\subfloat[]{
			\includegraphics[trim={1.2cm 0 0 1.0cm},clip, width=0.435\textwidth]{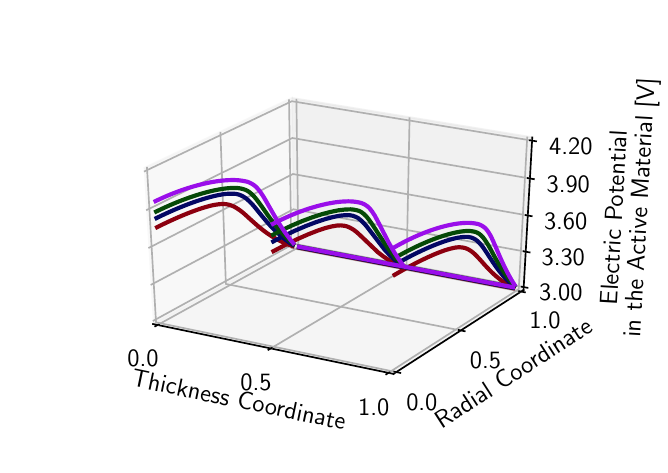}
			\label{img:epssec_phis}}\\
		\caption{(a) Normalized concentration in the active material and (b) electric potential in the solid phase at the end of a 5C discharge.}
		\label{img:epssec2}
	\end{center}
\end{figure*}

The first step to understand why a high secondary particle porosity yielded a lower rate capability lay in locating the capacity loss within the electrode structure. \ref{img:epssec_cs} displays the concentration profile in the active material at the end of 5C discharge over the electrode thickness and the secondary particle radius. Here, the normalized thickness coordinate runs from 0 at the separator to 1 at the current collector. For the normalized radial coordinate of the secondary particles at the respective thickness position, 0 denotes the particle center and 1 denotes the particle surface. Irrespective of the position within the electrode, the concentration in the active material decreased towards the center of the secondary particles. The concentration gradient intensified if the secondary particles were more porous. Corresponding to the concentration gradient, a gradient in electric potential as shown in \ref{img:epssec_phis} arose within the active material close to the surface of the secondary particles. Again, the gradient aggravated with increasing porosity of the secondary particles. From this we deduced that the effective electronic conductivity within the secondary particles limited the cell performance. The low electronic conductivity of NMC111 caused a poor exploitation of the active material close to the center of secondary particles. As the secondary particle porosity increased, the effective electronic conductivity further declined. In consequence, low secondary particle porosities outperformed high porosities. 

\begin{figure}[htb]
\begin{center}
\includegraphics[width=0.5\textwidth]{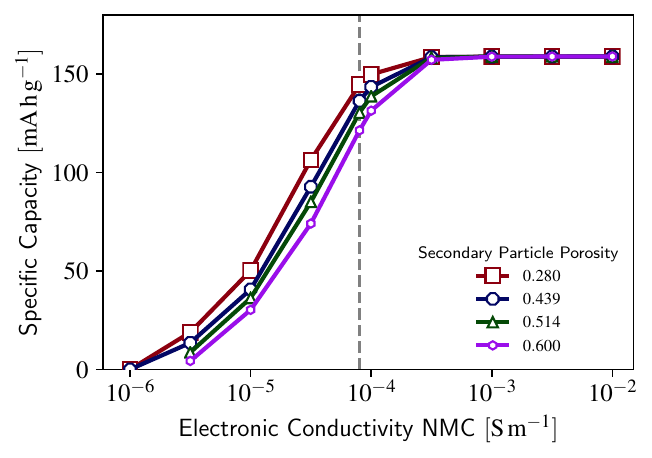}
\caption{Influence of electronic conductivity within the secondary particles on the cell performance at 5C.}
\label{img:epssec_sigma}
\end{center}
\end{figure}

A study of cells with varying electronic conductivity of the active material further validated our finding. Amin et al. \cite{amin_characterization_2016} found that the electronic conductivity of NMC111 varies between $10^{-6}~\si{\siemens\per\meter}$ and $10^{0}~\si{\siemens\per\meter}$ depending on lithiation and temperature. Furthermore, results from Kuo et al. \cite{kuo_contact_2022} suggested that measuring the electronic conductivity of NMC111 is difficult and includes uncertainties. For this reason, we chose a range of $10^{-6}~\si{\siemens\per\meter}$ to $10^{-2}~\si{\siemens\per\meter}$ for studying the qualitative influence of electronic conductivity on the cell performance. As apparent in \ref{img:epssec_sigma}, the cell performance scaled with the electronic conductivity in the range between $10^{-5.5}~\si{\siemens\per\meter}$ and $10^{-4}~\si{\siemens\per\meter}$. Above this range, the cell performance reached a limit and other physical processes in the electrode became restrictive. Below $10^{-5}~\si{\siemens\per\meter}$, the specific capacity at 5C declined rapidly. Already at $10^{-6}~\si{\siemens\per\meter}$, the electrode ceased to deliver any capacity. Raising the electronic conductivity in the secondary particles could enhance the rate capability of the cell, which proves that this material property is crucial for hierarchically structured electrodes. If the electronic conductivity of the active material is poor, an enhancement of the overall electronic conductivity in the secondary particle, e.g. by carbon coating of the primary particles, is essential to ensure a good cell performance \cite{lee_low-temperature_2017, phattharasupakun_core-shell_2021}. Our conclusions agreed with Wagner et al. \cite{wagner_hierarchical_2020} , Lueth et al. \cite{lueth_agglomerate_2015}, and Birkholz et al. \cite{birkholz_electrochemical_2021}, who identified the electronic conductivity as a main limit for the cell performance depending on the electrode morphology.

\begin{figure*}[htb]
	\begin{center}
		\subfloat[]{
			\includegraphics[width=0.5\textwidth]{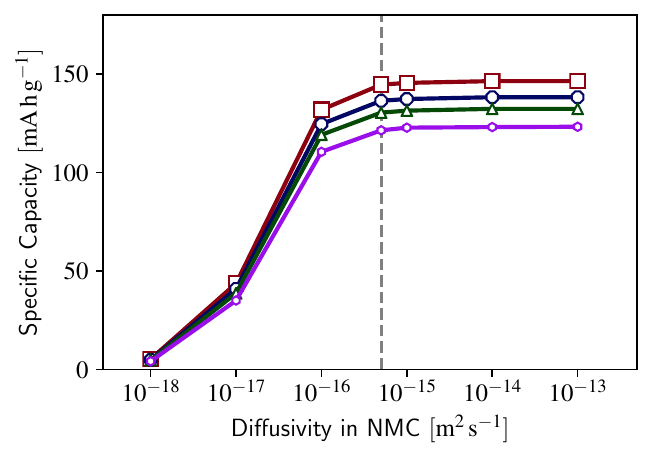}
			\label{img:epssec_Ds}}
		\subfloat[]{
			\includegraphics[trim={1.55cm 0 0 0},clip, width=0.435\textwidth]{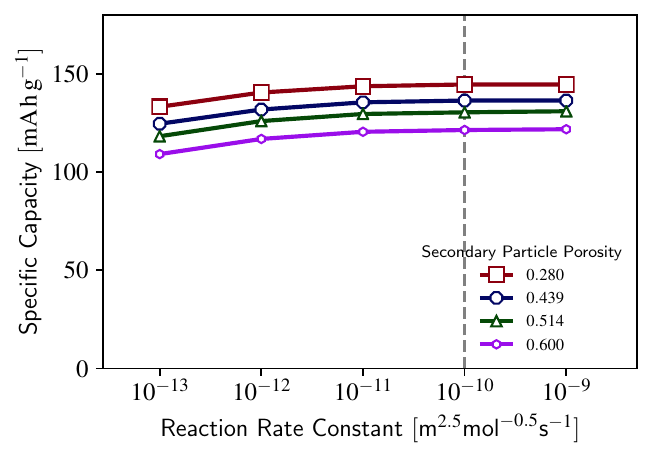}
			\label{img:epssec_k0}}\\
		\caption{Influence of (a) diffusivity in the active material and (b) reaction rate on the cell performance at 5C.}
		\label{img:epssec_mat}
	\end{center}
\end{figure*}

Other material properties had less influence on the relationship between secondary particle porosity and specific capacity. The specific capacity at 5C hardly changed for diffusivities between $10^{-16}~\si{\meter\squared\per\second}$ and $10^{-13}~\si{\meter\squared\per\second}$, see \ref{img:epssec_Ds}. A diffusivity below $10^{-16}~\si{\meter\squared\per\second}$ caused the difference in cell performance at 5C to decrease. This went hand in hand with an overall drastic decrease of specific capacity for this regime, until it completely faded at $10^{-18}~\si{\meter\squared\per\second}$ for all values of secondary particle porosity. When altering the reaction rate constant, the difference between the cell performance remained the same, see \ref{img:epssec_k0}. In summary, a change in diffusivity of the active material or in reaction rate constant hardly affected how the secondary particle porosity related to the cell performance.

\subsection{Dependence of electronic conductivity on the secondary particle size}

\begin{figure*}[htb]
	\begin{center}
		\subfloat[]{
			\includegraphics[width=0.5\textwidth]{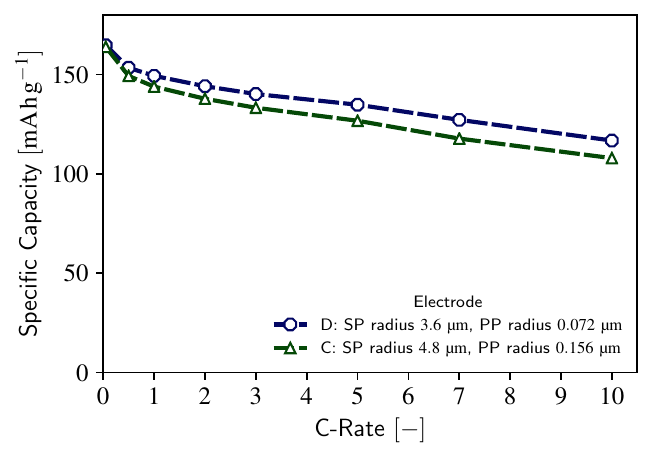}
			\label{img:rsec_exp}}
		\subfloat[]{
			\includegraphics[trim={1.55cm 0 0 0},clip, width=0.435\textwidth]{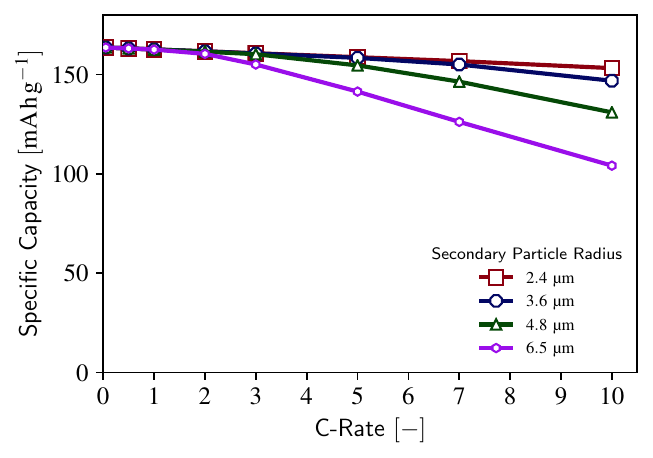}
			\label{img:rsec_sim}}\\
		\caption{(a) Experimental rate capability of electrodes C and D with differing secondary and primary particle size, see \ref{tab:powders} and \ref{tab:elec}. (b) Modeled rate capability depending on secondary particle size.}
		\label{img:rsec}
	\end{center}
\end{figure*}

In the previous section, we identified the electronic conductivity within the secondary particles as a limiting factor for the rate capability of hierarchically structured electrodes. Naturally, the question arose as to what extent the secondary particle size affects the performance of hierarchically structured electrodes. The secondary particle size determines the path lengths of electronic and ionic transport within the secondary particle. The hierarchically structured half-cell model includes the secondary particle size at the secondary particle level via the range of the radial coordinate $r^{(\textrm{II})}$, see Supporting Information Eq.~14-16. In addition to a varying secondary particle porosity, the secondary particles of electrodes A and B from the previous section showed a difference in size. Electrode B with smaller and less porous secondary particles outperformed electrode A in terms of rate capability, see \ref{img:epssec_exp}. In addition, we considered electrodes C and D, which differed both in secondary and primary particle size. \ref{img:rsec_exp} showed a higher rate capability of electrode D, consisting of smaller secondary and primary particles. Again, the hierarchically structured half-cell model could elucidate the impact of each single property. This section contains modeled results of electrodes with different secondary particle size to study its impact, while all other properties corresponded to the reference electrode. As displayed in \ref{img:rsec_sim}, the rate capability decreased with increasing secondary particle size. This indicated that the difference in secondary particle size could explain the difference in performance between electrodes A and B as well as between electrodes C and D to some extent. In electrodes A and B, the effects of secondary particle porosity and secondary particle size combined.

\begin{figure*}[htb]
	\begin{center}
		\subfloat[]{
			\includegraphics[trim={1.2cm 0 0 1.0cm},clip, width=0.5\textwidth]{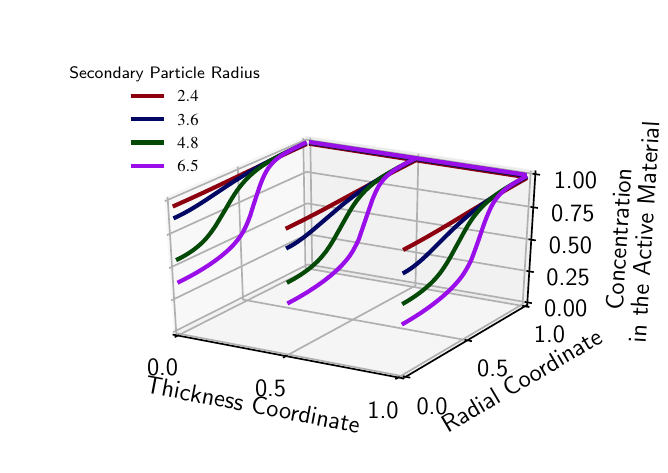}
			\label{img:rsec_cs}}
		\subfloat[]{
			\includegraphics[trim={1.2cm 0 0 1.0cm},clip, width=0.435\textwidth]{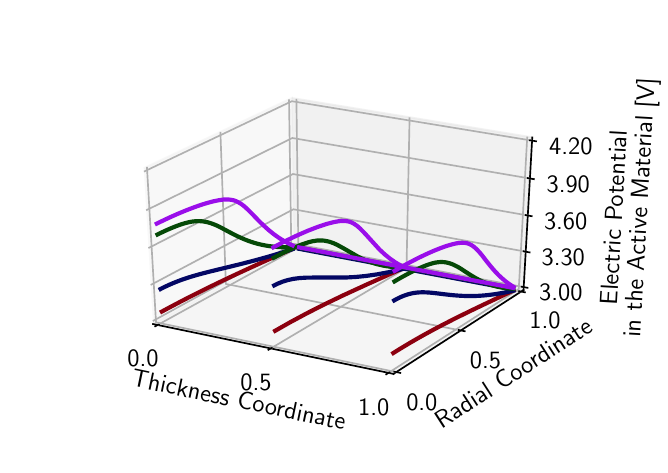}
			\label{img:rsec_phis}}\\
		\caption{(a) Normalized concentration in the active material and (b) electric potential in the solid phase at the end of a 5C discharge.}
		\label{img:rsec_}
	\end{center}
\end{figure*}

The loss in rate capability for larger secondary particles originated from an increased concentration drop over the secondary particle radius, see \ref{img:rsec_cs}. In the centers of large secondary particles, the active material was used to less of its capacity compared to small particles. The reason again lay in the low electronic conductivity within the secondary particles, as shown by steeper drops in electric potential with increasing secondary particle size displayed in \ref{img:rsec_phis}. The long electronic transport paths to the center of large secondary particles impairs rate capability if the electronic conductivity of the active material is low.

\begin{figure}[htb]
	\begin{center}
		\includegraphics[width=0.5\textwidth]{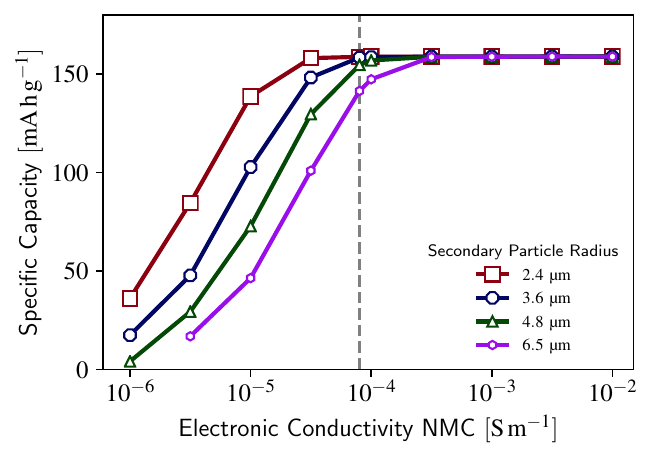}
		\caption{Influence of electronic conductivity within the secondary particles on the cell performance at 5C.}
		\label{img:rsec_sigma}
	\end{center}
\end{figure}

Investigating electrodes with varying electronic conductivity in the secondary particles confirmed its relation with the secondary particle size, see \ref{img:rsec_sigma}. If the electronic conductivity lay between $10^{-6}~\si{\siemens\per\meter}$ and $10^{-4}~\si{\siemens\per\meter}$, a larger secondary particle size significantly reduced the cell performance at high rates. Increasing the electronic conductivity raised the specific capacity at 5C from less than $40~\si{\milli\ampere\hour\per\gram}$ at the lowest electronic conductivity to a maximum of $159~\si{\milli\ampere\hour\per\gram}$ at $10^{-3}~\si{\siemens\per\meter}$ or higher. For small secondary particles up to $4.8~\si{\micro\meter}$ radius, an electronic conductivity of $10^{-4}~\si{\siemens\per\meter}$ sufficed to ensure an optimal specific capacity. Below this electronic conductivity, the size of the secondary particles significantly governed the rate capability. Our findings deviated from Wagner et al. \cite{wagner_hierarchical_2020}, who reported an improved rate capability for large secondary particles. However, the compared electrodes not only differed in secondary particle size, but also in secondary particle porosity and electrode porosity. Possibly, the influence of these properties surpassed the influence of the secondary particle size.

\begin{figure*}[htb]
	\begin{center}
		\subfloat[]{
			\includegraphics[width=0.5\textwidth]{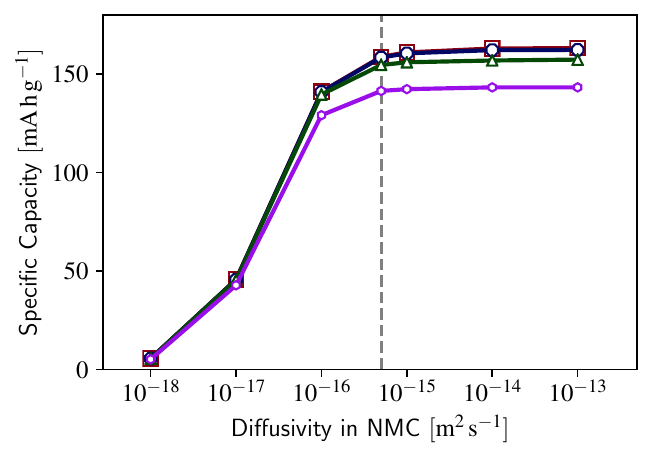}
			\label{img:rsec_Ds}}
		\subfloat[]{
			\includegraphics[trim={1.55cm 0 0 0},clip, width=0.435\textwidth]{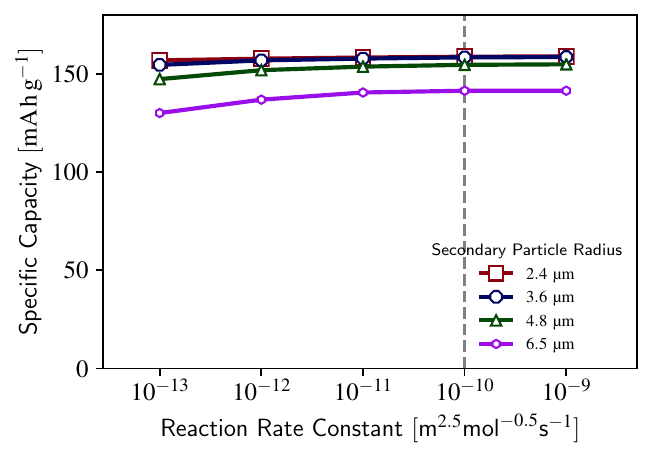}
			\label{img:rsec_k0}}\\
		\caption{Influence of (a) diffusivity in the active material and (b) reaction rate on the cell performance at 5C.}
		\label{img:rsec_mat}
	\end{center}
\end{figure*}

Similar to the secondary particle porosity, the secondary particle size played a role if the diffusivity in the active material lay above $10^{-16}~\si{\meter\squared\per\second}$, see \ref{img:rsec_Ds}. For lower diffusivities, the specific capacity at 5C declined. At the same time, the behavior of electrodes with different secondary particle size assimilated. Now the diffusion in the active material limited the cell performance. Therefore the effects of the low electronic conductivity in the secondary particle lost their relevance. \ref{img:rsec_k0} illustrated that the reaction rate constant barely affected the performance of electrodes with different secondary particle size.

\subsection{Limitations due to primary particle size and solid diffusion}

\begin{figure}[htb]
	\begin{center}
		\includegraphics[width=0.5\textwidth]{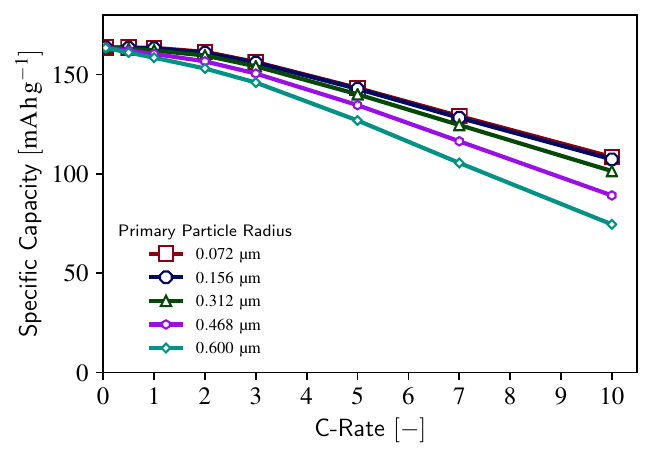}
		\caption{Modeled rate capability depending on primary particle size.}
		\label{img:rprim}
	\end{center}
\end{figure}

In the previous section we showed that a difference in secondary particle size caused the rate capability of electrodes C and D to diverge. Since these electrodes also differed in primary particle size, we now studied its influence with the hierarchically structured half-cell model. The primary particle size determines the length of diffusion paths within the active material, i.e. the range of the radial coordinate $r^{(\textrm{I})}$ (Supporting Information Eq.23). Note that after studying the electrochemically active surface area separately in section \ref{ch:eas_k0}, in this section the electrochemically active surface area changed according to the primary particle size. The model shown in \ref{img:rprim} yielded a lower rate capability for large primary particles. However, the model predicted barely any difference between primary particle radii of $0.072~\si{\micro\meter}$ and $0.156~\si{\micro\meter}$, which corresponded to electrodes C and D, see \ref{img:rsec_exp}. We concluded that the diverging secondary particle size accounted for almost all of the performance alteration between the two electrodes, while the primary particle size only contributed a small additional deviation. The primary particle size  plays a role in determining the rate capability of hierarchically structured electrodes if the radius exceeds around $0.300~\si{\micro\meter}$.

\begin{figure*}[htb]
	\begin{center}
		\subfloat[]{
			\includegraphics[trim={1.2cm 0 0 1.0cm},clip, width=0.5\textwidth]{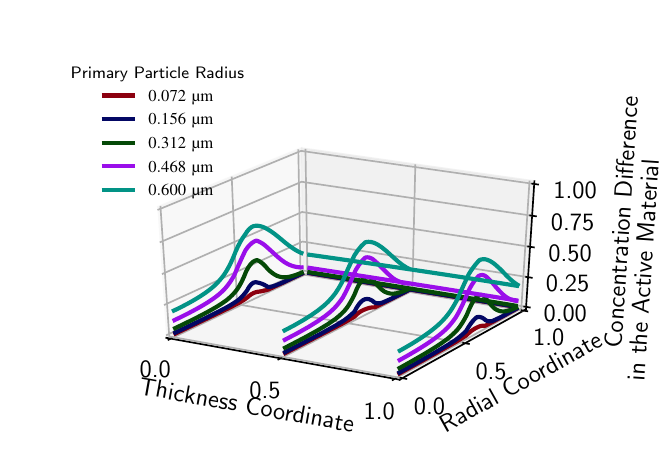}
			\label{img:rprim_deltacs}}
		\subfloat[]{
			\includegraphics[width=0.435\textwidth]{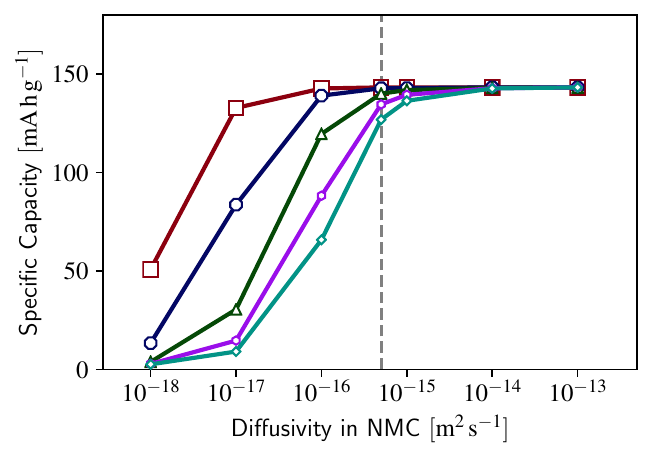}
			\label{img:rprim_Ds}}\\
		\caption{(a) Difference in normalized concentration in the active material between primary particle surfaces and centers at the end of a 5C discharge. The magnitude in concentration difference indicates the magnitude in concentration drop from the surface to the center of primary particles at each position in the electrode. (b) Influence of diffusivity in the active material on the cell performance at 5C.}
		\label{img:rprim_}
	\end{center}
\end{figure*}

The loss in rate capability due to a large primary particle size could be attributed to concentration drops inside the primary particles. \ref{img:rprim_deltacs} depicts the difference in concentration between primary particle surfaces and centers. The increased difference for large primary particles in one specific region of the secondary particles indicated larger concentration drops due to diffusion limitations in the active material. This region lay close enough to the secondary particle surface such that it did not suffer from the poor electronic conductivity in the secondary particles. In the center of the secondary particles, the electronic conductivity limited the intercalation, so that the concentration was low, independent of the primary particle size.

Studies of electrodes with different diffusivity of the active material in \ref{img:rprim_Ds} further confirmed our explanation. Electrodes containing small primary particles with a radius of $0.072~\si{\micro\meter}$ remained close to the maximum performance  until a diffusivity as low as $10^{-17}~\si{\meter\squared\per\second}$. With increasing primary particle size, the specific capacity already started dropping at higher diffusivity. At $10^{-18}~\si{\meter\squared\per\second}$, electrodes with primary particle radii below $0.156~\si{\micro\meter}$ failed completely. From this it followed that the diffusion in the active material can restrain the cell performance of hierarchically structured electrodes. Small primary particles therefore achieve the best exploitation of the active material and take full advantage of the reduced diffusion paths in hierarchically structured electrodes. In contrast to our conclusion, Birkholz et al. \cite{birkholz_electrochemical_2021} found the cell performance to be entirely independent of the diffusivity. In comparison, we used a refined value for the diffusivity in the active material and considered diffusion paths longer than the primary particle radius. With our more realistic approach we could show that an impact of the primary particle size is possible even in hierarchically structured electrodes. 

\begin{figure*}[htb]
	\begin{center}
		\subfloat[]{
			\includegraphics[width=0.5\textwidth]{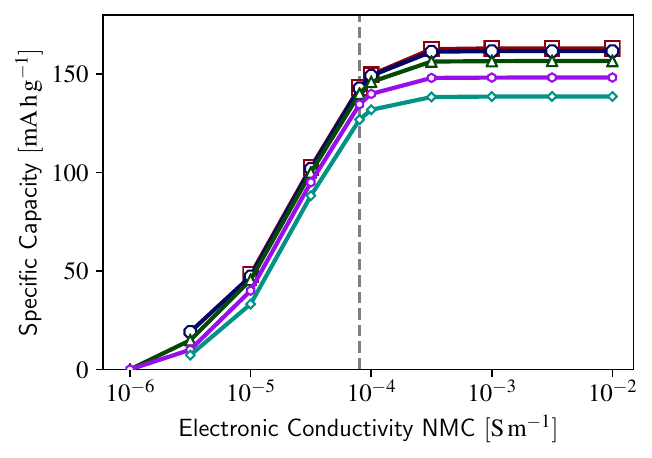}
			\label{img:rprim_sigma}}
		\subfloat[]{
			\includegraphics[trim={1.55cm 0 0 0},clip, width=0.435\textwidth]{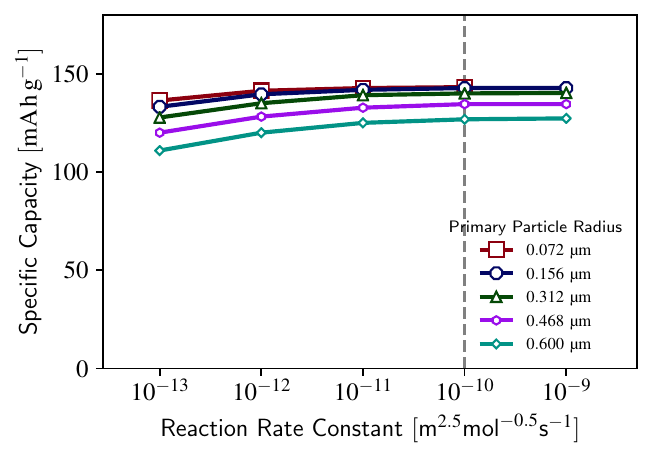}
			\label{img:rprim_k0}}\\
		\caption{Influence of (a) electronic conductivity of the active material and (b) reaction rate on the cell performance at 5C.}
		\label{img:rprim_mat}
	\end{center}
\end{figure*}

The effect of the primary particle size intensified in magnitude after increasing the electronic conductivity of the active material in \ref{img:rprim_sigma}. At $10^{-3}~\si{\siemens\per\meter}$ and higher, the absolute value of the specific capacity at 5C varied more strongly between different particle sizes than for the reference electrode with $8 \cdot 10^{-5}~\si{\siemens\per\meter}$. The absolute variation decreased for lower electronic conductivities while the overall cell performance diminished. At $10^{-6}~\si{\siemens\per\meter}$, the cell performance almost completely faded. As explained in the section \ref{ch:res_sigma}, a low electronic conductivity of the active material can restrict the cell performance. This agrees with our finding that the primary particle size plays a more important role if the electronic conductivity of the active material is high, since this allows diffusion to become more prominent as limiting factor. Based on experimental investigations, Wagner et al. \cite{wagner_hierarchical_2020} also drew the conclusion that large primary particles can limit the cell performance. At low reaction rates, the impact of primary particle size slightly increased as \ref{img:rprim_k0} indicated. If the maximum reaction rate droped below $10^{-11}~\text{m}^{2.5}\text{mol}^{-0.5}\text{s}^{-1}$, the specific capacity at 5C differed more significantly. Large primary particles imply a lower electrochemically active surface area for the same active material content of the electrode. Therefore, the electrochemical reaction slowed down, see section \ref{ch:eas_k0}.

\subsection{Impact of ionic transport depending on electrode thickness} 

\begin{figure*}[htb]
	\begin{center}
		\subfloat[]{
			\includegraphics[width=0.5\textwidth]{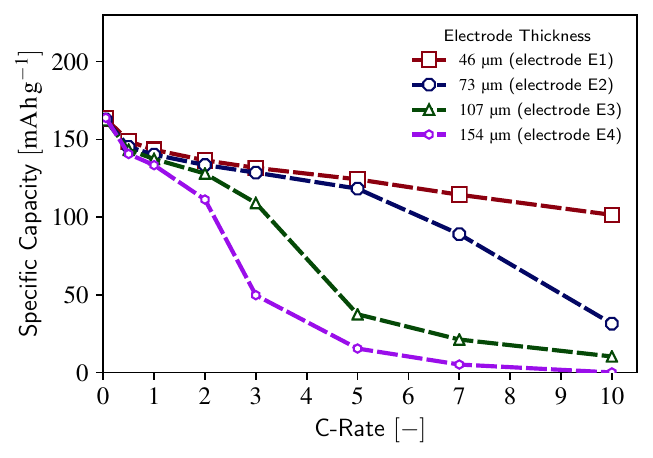}
			\label{img:Lpos_exp}}
		\subfloat[]{
			\includegraphics[trim={1.55cm 0 0 0},clip, width=0.435\textwidth]{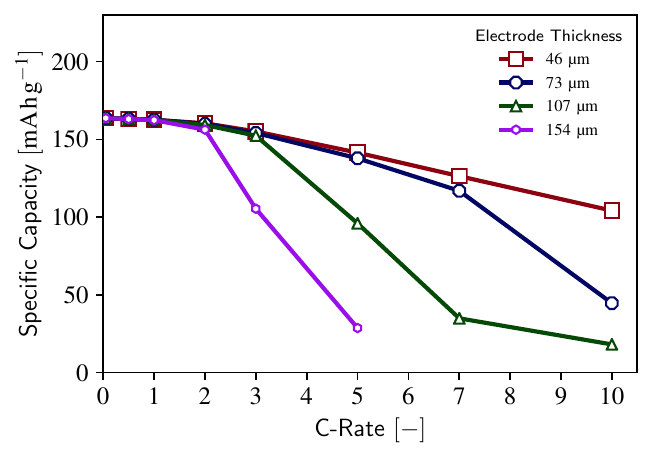}
			\label{img:Lpos_sim}}\\
		\caption{(a) Experimental (electrode E1-4, see \ref{tab:elec}) and (b) modeled rate capability depending on electrode thickness.}
		\label{img:Lpos}
	\end{center}
\end{figure*}

Finally, we investigated the electrode thickness of hierarchically structured electrodes. The hierarchically structured half-cell model incorporates the electrode thickness at the cell level (Supporting Information Eq.~1-3) by the range of the spatial coordinate $x$. The modeling studies depicted in \ref{img:Lpos_sim} displayed the same qualitative behavior as the experiment in \ref{img:Lpos_exp}. For progressively thick electrodes the rate capability decreased. Electrodes of $107~\si{\micro\meter}$ and thicker experienced a drastic drop in specific capacity within a small range of discharge rates before the decrease continued moderately above 7C. 

\begin{figure*}[htb]
\begin{center}
	\subfloat[]{
		\includegraphics[trim={1.2cm 0 0 1.0cm},clip, width=0.5\textwidth]{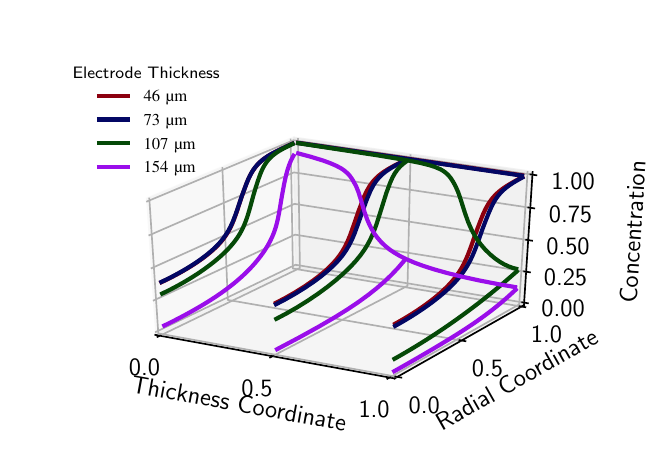}
		\label{img:Lpos_cs}}
	\subfloat[]{
		\includegraphics[trim={1.2cm 0 0 1.0cm},clip, width=0.435\textwidth]{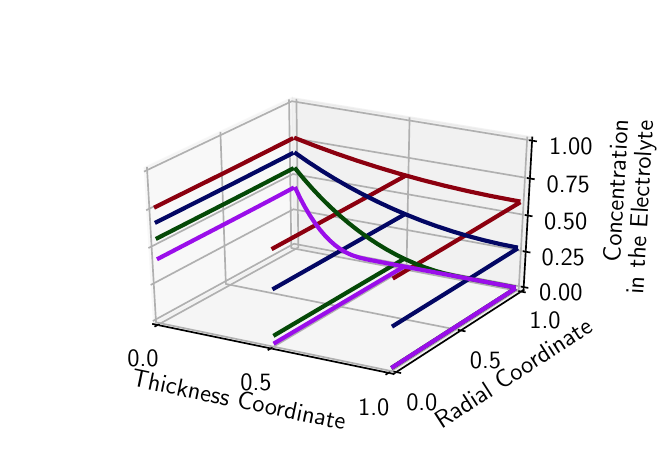}
		\label{img:Lpos_ce}}\\
	\caption{Normalized concentration (a) in the active material and (b) in the electrolyte at the end of a 5C discharge.}
	\label{img:Lpos_}
\end{center}
\end{figure*}

We further investigated the reason for the changing rate capability of hierarchically structured electrodes of different thickness. For thin electrodes up to $73~\si{\micro\meter}$, \ref{img:Lpos_cs} revealed a general slight decrease of concentration in the active material within the secondary particles. Furthermore, at a thickness of $107~\si{\micro\meter}$ a steep drop in concentration close to the current collector occured. Electrodes which were even thicker reached the low level of concentration across a larger portion of the thickness. As \ref{img:Lpos_ce} showed, the concentration drop in the active material related to a concentration drop in the electrolyte over the electrode thickness. This drop steepened with increasing electrode thickness. A lower concentration in the electrolyte caused the electrochemical reaction to run slower, so less concentration entered the active material.  Electrodes of $107~\si{\micro\meter}$ or thicker even experienced a complete depletion of the electrolyte in regions close to the current collector. When the electrolyte depleted, the electrochemical reaction stopped. This explained the drastically reduced concentration in the active material of these regions at the end of discharge. From our results, we deduce that the ionic transport in the electrolyte is so low that this process  compromises the cell performance in thick electrodes.

\begin{figure}[htb]
	\begin{center}
		\includegraphics[width=0.5\textwidth]{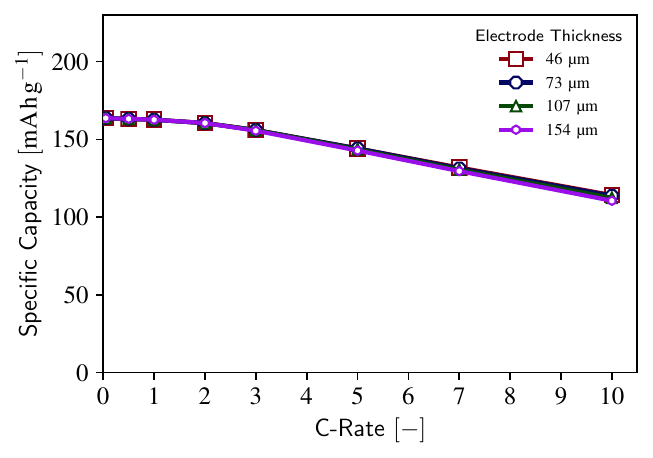}
		\caption{Rate capability when the conductivity and the diffusivity of the electrolyte are artificially increased by a factor of 10.}
		\label{img:Lpos_x10}
	\end{center}
\end{figure}

For additional proof of our finding, we increased the conductivity and the diffusivity of the electrolyte by a factor of 10. This resulted in an almost identical rate capability of the cells independent of the electrode thickness, see \ref{img:Lpos_x10}. Therefore, we confirmed that concentration drops in the electrolyte reduced the performance of thick electrodes. Classical electrodes with dense secondary active material particles experience the same effect \cite{du_understanding_2017, kuang_thick_2019}. Note that for even thicker electrodes or higher discharge rates the electronic transport through the solid phase of the electrode, which is mainly determined by conductive additives, may become limiting \cite{xu_modeling_2019}.

\section{Conclusion}
\label{conclusion}

Our goal was to elucidate the relationship between the morphology of hierarchically structured electrodes, material properties, and the rate capability of the cell.  The hierarchically structured half-cell model provides the possibility to independently study the influence of single morphology properties and material properties, which lies beyond experimental capabilities. In addition, the model predicts local state changes and thereby tracks the electrochemical processes within the cell. 

With our modeling study, we outlined the interplay between electrochemical processes within the novel concept of hierarchically structured electrodes. Electronic transport in the secondary particles is the dominant process which determines the rate capability. In case of a low electronic conductivity of the active material and no additional conductive aids inside the secondary particle, the transport of electrons to the primary particle surfaces where the electrochemical reaction takes place is insufficient. Despite the main transport taking place via conductive additives surrounding the secondary particles and comparatively short electronic transport paths within the secondary particles, the latter  play a dominant role since the electronic conductivity of additives and active material differ several orders of magnitude. A low secondary particle porosity and a small secondary particle size reduce the limitations due to electronic conductivity.

Hierarchically structured electrodes rely on the small size of the primary particles for short diffusion paths. Still, the diffusion within large primary particles may limit rate capability if the diffusivity in the active material is low. Compared to conventional electrode structures the limitation is mitigated. Electrode thickness affects hierarchically structured electrodes in a similar way as classical granular electrodes consisting of dense secondary particles. Namely, with increasing thickness depletion of the electrolyte reduces the rate capability of the cell. While transport proved to be limiting, the cell performance is insensitive to the experimentally achievable variation of electrochemically active surface area for hierarchically structured electrodes contrary to previous anticipation. This insight was enabled by separating effects of the electrochemically active surface area from primary particle size in simulations, while it is nearly impossible to vary these properties independent of each other when processing electrodes.

From our study, we deduced the following recommendations for the design of hierarchically structured electrodes. To increase the rate capability, a good electronic conductivity within the secondary particles of at least $10^{-4}~\si{\siemens\per\meter}$ is crucial. In addition, if the diffusivity in the active material lies below $10^{-15}~\si{\meter\squared\per\second}$, a small primary particle radius below $100~\si{\nano\meter}$ fully exploits the advantage of short diffusion paths in hierarchically structured electrodes. Ultrafast applications require thin electrodes, at most $70~\si{\micro\meter}$, or electrolytes with a high ion mobility. Our findings present a guideline for the design of high-performing electrodes of hierarchical structure. Conclusions for hierarchically structured electrodes were derived for NMC111 electrodes. But the general relationships between morphology, material properties, and rate capability likely apply to other systems as well, in particular for active material with poor electronic conductivity.

\clearpage

\section{Acknowledgements}

This work contributes to the research performed at CELEST (Center for Electrochemical Energy Storage Ulm-Karlsruhe) and was funded by the German Research Foundation (DFG) under Project ID 390874152 (POLiS Cluster of Excellence, EXC 2154). This work has been supported by the German Federal Ministry for Economic Affairs and Climate Action (BMWK) under reference number 03ETE039J. The responsibility for the content of this publication lies with the authors.

\section{Conflict of Interest}

The authors declare no conflict of interest. The funders had no role in the design of the study; in the collection, analyses, or interpretation of data; in the writing of the manuscript; or in the decision to publish the results.

\bibliographystyle{unsrt}
\bibliography{MyLibrarySelection.bib}

\begin{thebibliography}{10}

\bibitem{chen_hierarchical_2016}
Zhen Chen, Jin Wang, Dongliang Chao, Tom Baikie, Linyi Bai, Shi Chen, Yanli
  Zhao, Tze~Chien Sum, Jianyi Lin, and Zexiang Shen.
\newblock Hierarchical {Porous} {LiNi} 1/3 {Co} 1/3 {Mn} 1/3 {O} 2
  {Nano}-/{Micro} {Spherical} {Cathode} {Material}: {Minimized} {Cation}
  {Mixing} and {Improved} {Li} + {Mobility} for {Enhanced} {Electrochemical}
  {Performance}.
\newblock {\em Sci Rep}, 6(1):1--10, May 2016.

\bibitem{mueller_effect_2021}
Marcus Mueller, Luca Schneider, Nicole Bohn, Joachim~R. Binder, and Werner
  Bauer.
\newblock Effect of {Nanostructured} and {Open}-{Porous} {Particle}
  {Morphology} on {Electrode} {Processing} and {Electrochemical} {Performance}
  of {Li}-{Ion} {Batteries}.
\newblock {\em ACS Appl. Energy Mater.}, 4(2):1993--2003, February 2021.

\bibitem{wagner_hierarchical_2020}
Amalia~Christina Wagner, Nicole Bohn, Holger Geßwein, Matthias Neumann, Markus
  Osenberg, Andre Hilger, Ingo Manke, Volker Schmidt, and Joachim~R. Binder.
\newblock Hierarchical {Structuring} of {NMC111}-{Cathode} {Materials} in
  {Lithium}-{Ion} {Batteries}: {An} {In}-{Depth} {Study} on the {Influence} of
  {Primary} and {Secondary} {Particle} {Sizes} on {Electrochemical}
  {Performance}.
\newblock {\em ACS Appl. Energy Mater.}, 3(12):12565--12574, December 2020.

\bibitem{neumann_data-driven_2023}
Matthias Neumann, Sven~E. Wetterauer, Markus Osenberg, André Hilger, Phillip
  Gräfensteiner, Amalia Wagner, Nicole Bohn, Joachim~R. Binder, Ingo Manke,
  Thomas Carraro, and Volker Schmidt.
\newblock A {Data}-{Driven} {Modeling} {Approach} to {Quantify} {Morphology}
  {Effects} on {Transport} {Properties} in {Nanostructured} {NMC} {Particles}.
\newblock preprint, SSRN, 2023.

\bibitem{wu_mechanical-electrochemical_2016}
Bin Wu and Wei Lu.
\newblock Mechanical-{Electrochemical} {Modeling} of {Agglomerate} {Particles}
  in {Lithium}-{Ion} {Battery} {Electrodes}.
\newblock {\em J. Electrochem. Soc.}, 163(14):A3131, November 2016.

\bibitem{cernak_three-dimensional_2021}
Susanne Cernak, Florian Schuerholz, Michael Kespe, and Hermann Nirschl.
\newblock Three-{Dimensional} {Numerical} {Simulations} on the {Effect} of
  {Particle} {Porosity} of
  {Lithium}-{Nickel}–{Manganese}–{Cobalt}–{Oxide} on the {Performance} of
  {Positive} {Lithium}-{Ion} {Battery} {Electrodes}.
\newblock {\em Energy Technol.}, 9(6):2000676, 2021.

\bibitem{lueth_agglomerate_2015}
S.~Lueth, U.~S. Sauter, and W.~G. Bessler.
\newblock An {Agglomerate} {Model} of {Lithium}-{Ion} {Battery} {Cathodes}.
\newblock {\em J. Electrochem. Soc.}, 163(2):A210, November 2015.

\bibitem{birkholz_electrochemical_2021}
Oleg Birkholz and Marc Kamlah.
\newblock Electrochemical {Modeling} of {Hierarchically} {Structured}
  {Lithium}-{Ion} {Battery} {Electrodes}.
\newblock {\em Energy Technol.}, 9(6):2000910, 2021.

\bibitem{dreizler_investigation_2018}
Andreas~M. Dreizler, Nicole Bohn, Holger Gesswein, Marcus Mueller, Joachim~R.
  Binder, Norbert Wagner, and K.~Andreas Friedrich.
\newblock Investigation of the {Influence} of {Nanostructured}
  {LiNi0}.{33Co0}.{33Mn0}.{33O2} {Lithium}-{Ion} {Battery} {Electrodes} on
  {Performance} and {Aging}.
\newblock {\em J. Electrochem. Soc.}, 165(2):A273, January 2018.

\bibitem{doyle_modeling_1993}
Marc Doyle, Thomas~F. Fuller, and John Newman.
\newblock Modeling of {Galvanostatic} {Charge} and {Discharge} of the
  {Lithium}/{Polymer}/{Insertion} {Cell}.
\newblock {\em J. Electrochem. Soc.}, 140(6):1526, June 1993.

\bibitem{fuller_simulation_1994}
Thomas~F. Fuller, Marc Doyle, and John Newman.
\newblock Simulation and {Optimization} of the {Dual} {Lithium} {Ion}
  {Insertion} {Cell}.
\newblock {\em J. Electrochem. Soc.}, 141(1):1--10, January 1994.

\bibitem{doyle_design_1995}
C.~M. Doyle.
\newblock Design and simulation of lithium rechargeable batteries.
\newblock Technical Report LBL-37650, Lawrence Berkeley National Lab. (LBNL),
  Berkeley, CA (United States), August 1995.

\bibitem{doyle_modeling_1995}
Marc Doyle and John Newman.
\newblock Modeling the performance of rechargeable lithium-based cells: design
  correlations for limiting cases.
\newblock {\em J. Power Sources}, 54(1):46--51, March 1995.

\bibitem{doyle_use_1995}
Marc Doyle and John Newman.
\newblock The use of mathematical modeling in the design of lithium/polymer
  battery systems.
\newblock {\em Electrochim. Acta}, 40(13):2191--2196, October 1995.

\bibitem{doyle_comparison_1996}
Marc Doyle, John Newman, Antoni~S. Gozdz, Caroline~N. Schmutz, and Jean-Marie
  Tarascon.
\newblock Comparison of {Modeling} {Predictions} with {Experimental} {Data}
  from {Plastic} {Lithium} {Ion} {Cells}.
\newblock {\em J. Electrochem. Soc.}, 143(6):1890, June 1996.

\bibitem{doyle_analysis_1997}
M.~DOYLE and J.~NEWMAN.
\newblock Analysis of capacity–rate data for lithium batteries using
  simplified models of the discharge process.
\newblock {\em J. Appl. Electrochem.}, 27(7):846--856, July 1997.

\bibitem{doyle_computer_2000}
Marc Doyle, Jeremy~P. Meyers, and John Newman.
\newblock Computer {Simulations} of the {Impedance} {Response} of {Lithium}
  {Rechargeable} {Batteries}.
\newblock {\em J. Electrochem. Soc.}, 147(1):99, January 2000.

\bibitem{doyle_computer_2003}
Marc Doyle and Yuris Fuentes.
\newblock Computer {Simulations} of a {Lithium}-{Ion} {Polymer} {Battery} and
  {Implications} for {Higher} {Capacity} {Next}-{Generation} {Battery}
  {Designs}.
\newblock {\em J. Electrochem. Soc.}, 150(6):A706, April 2003.

\bibitem{newman_porous-electrode_1975}
John Newman and William Tiedemann.
\newblock Porous-electrode theory with battery applications.
\newblock {\em AIChE Journal}, 21(1):25--41, 1975.

\bibitem{thomas_mathematical_2002}
Karen~E. Thomas, John Newman, and Robert~M. Darling.
\newblock Mathematical {Modeling} of {Lithium} {Batteries}.
\newblock In Walter~A. van Schalkwijk and Bruno Scrosati, editors, {\em
  Advances in {Lithium}-{Ion} {Batteries}}, pages 345--392. Springer US,
  Boston, MA, 2002.

\bibitem{newman_electrochemical_2012}
John Newman and Karen~E. Thomas-Alyea.
\newblock {\em Electrochemical {Systems}}.
\newblock John Wiley \& Sons, November 2012.
\newblock Google-Books-ID: eyj4MRa7vLAC.

\bibitem{schmidt_understanding_2021}
Adrian Schmidt, Elvedin Ramani, Thomas Carraro, Jochen Joos, André Weber, Marc
  Kamlah, and Ellen Ivers-Tiffée.
\newblock Understanding {Deviations} between {Spatially} {Resolved} and
  {Homogenized} {Cathode} {Models} of {Lithium}-{Ion} {Batteries}.
\newblock {\em Energy Technol.}, 9(6):2000881, 2021.

\bibitem{schneider_transport_2022}
Luca Schneider, Julian Klemens, Eike~Christian Herbst, Marcus Müller, Philip
  Scharfer, Wilhelm Schabel, Werner Bauer, and Helmut Ehrenberg.
\newblock Transport {Properties} in {Electrodes} for {Lithium}-{Ion}
  {Batteries}: {Comparison} of {Compact} versus {Porous} {NCM} {Particles}.
\newblock {\em J. Electrochem. Soc.}, 169(10):100553, October 2022.

\bibitem{amin_characterization_2016}
Ruhul Amin and Yet-Ming Chiang.
\newblock Characterization of {Electronic} and {Ionic} {Transport} in {Li}
  $_{\textrm{1-}}$ \textit{ $_{\textrm{x}}$ } {Ni} $_{\textrm{0.33}}$ {Mn}
  $_{\textrm{0.33}}$ {Co} $_{\textrm{0.33}}$ {O} $_{\textrm{2}}$ ({NMC}
  $_{\textrm{333}}$ ) and {Li} $_{\textrm{1-}}$ \textit{ $_{\textrm{x}}$ } {Ni}
  $_{\textrm{0.50}}$ {Mn} $_{\textrm{0.20}}$ {Co} $_{\textrm{0.30}}$ {O}
  $_{\textrm{2}}$ ({NMC} $_{\textrm{523}}$ ) as a {Function} of {Li} {Content}.
\newblock {\em J. Electrochem. Soc.}, 163(8):A1512--A1517, 2016.

\bibitem{kuo_contact_2022}
Jimmy~Jiahong Kuo, Stephen~Dongmin Kang, and William~C. Chueh.
\newblock Contact {Resistance} of {Carbon}–{Lix}({Ni},{Mn},{Co}){O2}
  {Interfaces}.
\newblock {\em Adv. Energy Mater.}, 12(31):2201114, 2022.

\bibitem{lee_low-temperature_2017}
Min-Joon Lee, Eunsol Lho, Peng Bai, Sujong Chae, Ju~Li, and Jaephil Cho.
\newblock Low-{Temperature} {Carbon} {Coating} of {Nanosized}
  {Li1}.{015Al0}.{06Mn1}.{925O4} and {High}-{Density} {Electrode} for
  {High}-{Power} {Li}-{Ion} {Batteries}.
\newblock {\em Nano Lett.}, 17(6):3744--3751, June 2017.

\bibitem{phattharasupakun_core-shell_2021}
Nutthaphon Phattharasupakun, Juthaporn Wutthiprom, Salatan Duangdangchote,
  Sangchai Sarawutanukul, Chanikarn Tomon, Farkfun Duriyasart, Suchakree
  Tubtimkuna, Chalita Aphirakaramwong, and Montree Sawangphruk.
\newblock Core-shell {Ni}-rich {NMC}-{Nanocarbon} cathode from scalable
  solvent-free mechanofusion for high-performance 18650 {Li}-ion batteries.
\newblock {\em Energy Stor. Mater.}, 36:485--495, April 2021.

\bibitem{du_understanding_2017}
Zhijia Du, D.~L. Wood, C.~Daniel, S.~Kalnaus, and Jianlin Li.
\newblock Understanding limiting factors in thick electrode performance as
  applied to high energy density {Li}-ion batteries.
\newblock {\em J Appl Electrochem}, 47(3):405--415, March 2017.

\bibitem{kuang_thick_2019}
Yudi Kuang, Chaoji Chen, Dylan Kirsch, and Liangbing Hu.
\newblock Thick {Electrode} {Batteries}: {Principles}, {Opportunities}, and
  {Challenges}.
\newblock {\em Adv. Energy Mater.}, 9(33):1901457, 2019.

\bibitem{xu_modeling_2019}
Meng Xu, Benjamin Reichman, and Xia Wang.
\newblock Modeling the effect of electrode thickness on the performance of
  lithium-ion batteries with experimental validation.
\newblock {\em Energy}, 186:115864, November 2019.

\end{thebibliography}

\clearpage



\section*{Supplementary material}

both at half-cell and secondary particle level. They include ionic mass transport (1 and 14), ionic charge transport (2 and 15), and electronic transport (3 and 16). In addition, the reduced species diffuses into the primary particles (23). Under galvanostatic discharge, a constant electronic current (13) enters the electrode at the current collector (8). Also, there is a constant influx of ions from the separator (9), which is electronically insulating (10). Ions cannot pass through the current collector (11 and 12). The influx of the reduced species at the primary particle surfaces (25) occurs according to the electrochemical reaction (6). All electrochemical quantities are continuous at the secondary particle surfaces (17, 18, and 19) and their gradients vanish in the secondary and primary particle centers due to spherical symmetry (20, 21, 22, and 24).

## 1.1 Quantities

[hp] $\bar{c}_e$: concentration in the electrolyte at half-cell level

$\bar{\phi}_e$: electrochemical potential in the electrolyte at half-cell level

$\bar{\phi}_s$: electrostatic potential in the solid phase of the electrode at half-cell level

$\bar{c}_e^{(II)}$: concentration in the electrolyte within the secondary particles

$\bar{\phi}_e^{(II)}$: electrochemical potential in the electrolyte within the secondary particles

$\bar{\phi}_s^{(II)}$: electrostatic potential in the solid phase of the secondary particles

$c_s^{(I)}$: concentration in the primary particles

$t$: time

$x$: through-thickness coordinate of the half-cell

$r^{(II)}$: radial coordinate of secondary particles

$r^{(I)}$: radial coordinate of primary particles

## 1.2 Half-cell level

$$\varepsilon_e \frac{\partial \bar{c}_e}{\partial t} = \frac{\partial}{\partial x}\left(D_{e,\text{eff}} \frac{\partial \bar{c}_e}{\partial x}\right) + \varepsilon_s \bar{j}_m \tag{1}$$

$$\frac{\partial}{\partial x}\left(\kappa_{e,\text{eff}} \frac{\partial \bar{\phi}_e}{\partial x} + \kappa_{D,\text{eff}} \frac{\partial \ln \bar{c}_e}{\partial x}\right) = -\varepsilon_s \bar{j}_{\text{ch}} \tag{2}$$

$$\frac{\partial}{\partial x}\left(\sigma_{\text{eff}} \frac{\partial \bar{\phi}_s}{\partial x}\right) = \varepsilon_s \bar{j}_{\text{ch}} \tag{3}$$

with

$$\varepsilon_s \bar{j}_m = \frac{3\varepsilon_s}{\left(R^{(II)}\right)^3} \int_0^{R^{(II)}} \left[ \left(r^{(II)}\right)^2 \varepsilon_e^{(II)} \frac{\partial \bar{c}_e^{(II)}}{\partial t} \right.$$
$$\left. - \left(r^{(II)}\right)^2 a_s^{(I)} (1-t_+) \bar{j} \right] dr^{(II)} \quad (4)$$

$$\varepsilon_s \bar{j}_{ch} = \frac{3\varepsilon_s}{\left(R^{(II)}\right)^3} \int_0^{R^{(II)}} \left(r^{(II)}\right)^2 a_s^{(I)} \bar{j} F \, dr^{(II)} \quad (5)$$

$$\bar{j} = k_0 \left(\bar{c}_e^{(II)}\right)^{1-\alpha} \left(\left[c_s^{(I)}\right]_{r^{(I)}=R^{(I)}}\right)^\alpha$$
$$\left(c_{s,max} - \left[c_s^{(I)}\right]_{r^{(I)}=R^{(I)}}\right)^{1-\alpha}$$
$$\left[\exp\left(\frac{(1-\alpha)F}{RT}\bar{\eta}^{(II)}\right) - \exp\left(\frac{-\alpha F}{RT}\bar{\eta}^{(II)}\right)\right] \quad (6)$$

$$\bar{\eta}^{(II)} = \bar{\phi}_s^{(II)} - \bar{\phi}_e^{(II)} - U_{OCV}\left(\left[c_s^{(I)}\right]_{r^{(I)}=R^{(I)}}\right) \quad (7)$$

Boundary conditions

$$\left[\sigma_{eff} \frac{\partial \bar{\phi}_s}{\partial x}\right]_{x=L^{sep}+L^{pos}} = -i_{app} \quad (8)$$

$$\left[D_{e,eff} \frac{\partial \bar{c}_e}{\partial x}\right]_{x=0} = i_{app}(1-t_+) \quad (9)$$

$$\left[\frac{\partial \bar{\phi}_s}{\partial x}\right]_{x=L^{sep}} = 0 \quad (10)$$

$$\left[\frac{\partial \bar{c}_e}{\partial x}\right]_{x=L^{sep}+L^{pos}} = 0 \quad (11)$$

$$\left[\kappa_{eff} \frac{\partial \bar{\phi}_e}{\partial x} + \kappa_{D,eff} \frac{\partial \ln \bar{c}_e}{\partial x}\right]_{x=L^{sep}+L^{pos}} = 0 \quad (12)$$

with

$$i_{app} = C \frac{L^{pos} \varepsilon_s^{(II)} \varepsilon_s F (c_{s,max} - c_{s,0})}{3600} \quad (13)$$

### 1.3 Secondary particle level

$$\varepsilon_e^{(II)} \frac{\partial \bar{c}_e^{(II)}}{\partial t} = \frac{1}{\left(r^{(II)}\right)^2} \frac{\partial}{\partial r^{(II)}} \left(\left(r^{(II)}\right)^2 D_{e,eff}^{(II)} \frac{\partial \bar{c}_e^{(II)}}{\partial r^{(II)}}\right)$$
$$+ a_s^{(I)}(1-t_+)\bar{j} \quad (14)$$

3

$$\frac{1}{\left(r^{(\mathrm{II})}\right)^2}\frac{\partial}{\partial r^{(\mathrm{II})}}\left(\left(r^{(\mathrm{II})}\right)^2 \kappa_{\mathrm{eff}}^{(\mathrm{II})}\frac{\partial \bar{\phi}_{\mathrm{e}}^{(\mathrm{II})}}{\partial r^{(\mathrm{II})}}\right.$$
$$\left. + \left(r^{(\mathrm{II})}\right)^2 \kappa_{\mathrm{D,eff}}^{(\mathrm{II})}\frac{\partial \ln \bar{c}_{\mathrm{e}}^{(\mathrm{II})}}{\partial r^{(\mathrm{II})}}\right) = -a_{\mathrm{s}}^{(\mathrm{I})}F\bar{j} \quad (15)$$

$$\frac{1}{\left(r^{(\mathrm{II})}\right)^2}\frac{\partial}{\partial r^{(\mathrm{II})}}\left(\left(r^{(\mathrm{II})}\right)^2 \sigma_{\mathrm{eff}}^{(\mathrm{II})}\frac{\partial \bar{\phi}_{\mathrm{s}}^{(\mathrm{II})}}{\partial r^{(\mathrm{II})}}\right) = a_{\mathrm{s}}^{(\mathrm{I})}F\bar{j} \quad (16)$$

Boundary conditions

$$\left[\bar{\phi}_{\mathrm{s}}^{(\mathrm{II})}\right]_{r^{(\mathrm{II})}=R^{(\mathrm{II})}} = \bar{\phi}_{\mathrm{s}} \quad (17)$$

$$\left[\bar{\phi}_{\mathrm{e}}^{(\mathrm{II})}\right]_{r^{(\mathrm{II})}=R^{(\mathrm{II})}} = \bar{\phi}_{\mathrm{e}} \quad (18)$$

$$\left[\bar{c}_{\mathrm{e}}^{(\mathrm{II})}\right]_{r^{(\mathrm{II})}=R^{(\mathrm{II})}} = \bar{c}_{\mathrm{e}} \quad (19)$$

$$\left[\frac{\partial \bar{\phi}_{\mathrm{s}}^{(\mathrm{II})}}{\partial r^{(\mathrm{II})}}\right]_{r^{(\mathrm{II})}=0} = 0 \quad (20)$$

$$\left[\frac{\partial \bar{\phi}_{\mathrm{e}}^{(\mathrm{II})}}{\partial r^{(\mathrm{II})}}\right]_{r^{(\mathrm{II})}=0} = 0 \quad (21)$$

$$\left[\frac{\partial \bar{c}_{\mathrm{e}}^{(\mathrm{II})}}{\partial r^{(\mathrm{II})}}\right]_{r^{(\mathrm{II})}=0} = 0 \quad (22)$$

### 1.4 Primary particle level

$$\frac{\partial c_{\mathrm{s}}^{(\mathrm{I})}}{\partial t} = \frac{1}{\left(r^{(\mathrm{I})}\right)^2}\frac{\partial}{\partial r^{(\mathrm{I})}}\left(D_{\mathrm{s}}^{(\mathrm{I})}\left(r^{(\mathrm{I})}\right)^2\frac{\partial c_{\mathrm{s}}^{(\mathrm{I})}}{\partial r^{(\mathrm{I})}}\right) \quad (23)$$

Boundary conditions

$$\left[\frac{\partial c_{\mathrm{s}}^{(\mathrm{I})}}{\partial r^{(\mathrm{I})}}\right]_{r^{(\mathrm{I})}=0} = 0 \quad (24)$$

$$\left[D_{\mathrm{s}}^{(\mathrm{I})}\frac{\partial c_{\mathrm{s}}^{(\mathrm{I})}}{\partial r^{(\mathrm{I})}}\right]_{r^{(\mathrm{I})}=R^{(\mathrm{I})}} = -\frac{\left(R^{(\mathrm{I})}/1.5\right)a_{\mathrm{s}}^{(\mathrm{I})}}{3\varepsilon_{\mathrm{s}}^{(\mathrm{II})}}\bar{j} \quad (25)$$

## 2 Properties of the E1 reference cell

### 2.1 Geometric and electrochemical properties

Functions within the model:

$$\varepsilon_{\mathrm{e}}^{\mathrm{sep}} = 1 - \frac{\rho^{\mathrm{sep}}}{\rho_{\mathrm{bsg}}} = 1 - \frac{\frac{w^{\mathrm{sep}}}{L^{\mathrm{sep}}}}{\rho_{\mathrm{bsg}}} = 1 - \frac{\frac{53\ \mathrm{g\,cm^{-2}}}{260\ \mathrm{\mu m}}}{2.23\ \mathrm{g\,cm^{-3}}} = 0.91 \quad (26)$$

$$\kappa_e = \Big(0.0798(1 + (T - 228))(1 - 1.22\sqrt{\hat{c}_e}$$
$$+ 0.509(1 - 0.004\exp(1000/T))\hat{c}_e)\hat{c}_e\,/$$
$$\big(1 + (0.00379\exp(1000/T))\hat{c}_e^4\big)\ \mathrm{S\,m^{-1}} \quad (27)$$

$$D_e = 1.47 \cdot 10^{-7} \cdot \exp\Big(1.33 \cdot \hat{c}_e - \frac{1.69 \cdot 10^3}{T}$$
$$-\frac{5.63 \cdot 10^2}{T} \cdot \hat{c}_e\Big)\ \mathrm{m^2\,s^{-1}} \quad (28)$$

$$\nu_{\mathrm{tdf}} = -5.58 + 7.17\hat{c}_e + 0.038T + 1.91\hat{c}_e^2$$
$$- 0.0665\hat{c}_e T - 0.0000508T^2 + 0.11\hat{c}_e^3$$
$$- 0.0061\hat{c}_e^2 T + 0.000151\hat{c}_e T^2 \quad (29)$$

$$c_{s,\max} = \frac{\mathrm{DOD}_{\max}Q_{\mathrm{spec}}\rho_s}{F} = \frac{Q_{\mathrm{spec}}\rho_s}{F} \quad (30)$$

$$c_{s,0} = \frac{\mathrm{DOD}_0 Q_{\mathrm{spec}}\rho_s}{F} \quad (31)$$

with

$$\hat{c}_e = \frac{\bar{c}_e^{\mathrm{sep}}}{\bar{c}_{e,0}}, \frac{\bar{c}_e^{\mathrm{pos}}}{\bar{c}_{e,0}}, \frac{\bar{c}_e^{(II)}}{\bar{c}_{e,0}}$$

$$\mathrm{DOD}_0 = \frac{Q_{\mathrm{spec}} - Q_{\mathrm{rev}}}{Q_{\mathrm{spec}}} \quad (32)$$

$$a_s^{(I)}[\mathrm{m^2\,g^{-1}}] = \frac{a_s^{(I)}[\mathrm{m^{-1}}]}{\rho_s} \quad (33)$$

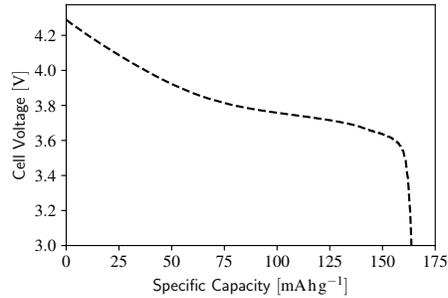

Figure 1: Open circuit voltage $U_{\mathrm{OCV}}$ (experimental discharge curve of the reference cell at C/20).

5

Table 1: Geometric properties of the reference cell.

| Parameter | Value | Description | Source |
|---|---|---|---|
| $L^{\text{sep}}$ | 260 µm | thickness of the separator | [2] |
| $L^{\text{pos}}$ | 46 µm | thickness of the positive electrode | |
| $R^{(\text{II})}$ | 6.5 µm | radius of the secondary particles | |
| $R^{(\text{I})}$ | $1.5 \cdot 0.255$ µm | representative length of diffusion paths within the primary particles | |
| $\varepsilon_{\text{e}}^{\text{sep}}$ | 0.91 | volume fraction of electrolyte in the separator | [2] |
| $\varepsilon_{\text{e}}^{\text{pos}}$ | $1 - \varepsilon_{\text{s}}^{\text{pos}} - \varepsilon_{\text{f}}^{\text{pos}}$ | volume fraction of electrolyte in the positive electrode | |
| $\varepsilon_{\text{s}}^{\text{pos}}$ | 0.5762 | volume fraction of secondary particles in the positive electrode | |
| $\varepsilon_{\text{f}}^{\text{pos}}$ | 0.1373 | volume fraction of additives in the positive electrode | |
| $\varepsilon_{\text{e}}^{(\text{II})}$ | $1 - \varepsilon_{\text{s}}^{(\text{II})}$ | volume fraction of electrolyte in the secondary particles = secondary particle porosity | |
| $\varepsilon_{\text{s}}^{(\text{II})}$ | 0.6480 | volume fraction of NMC111 in the secondary particles | |
| $a_{\text{s}}^{(\text{I})}$ | $3\dfrac{\varepsilon_{\text{s}}^{(\text{II})}}{R^{(\text{I})}/1.5}$ | electrochemically active surface area of the electrode | |

Table 2: Electrochemical properties of the reference cell.

| Parameter | Value | Description | Source |
|---|---|---|---|
| $\sigma$ | $100\ \text{S}\,\text{m}^{-1}$ | electronic conductivity of the carbon-binder domain | [3] |
| $\sigma^{(\text{II})}$ | $8\cdot 10^{-5}\ \text{S}\,\text{m}^{-1}$ | electronic conductivity of NMC111 | [4], [a] |
| $\kappa_\text{e}$ | Eq. (27) | conductivity of the electrolyte | [5] |
| $\kappa_\text{D}$ | $-\dfrac{2RT}{F}\kappa_\text{e}\nu_\text{tdf}(1-t_+)$ | diffusional conductivity of the electrolyte | [5] |
| $D_\text{e}$ | Eq. (28) | diffusivity in the electrolyte | [5] |
| $D_\text{s}$ | $5\cdot 10^{-16}\ \text{m}^2\,\text{s}^{-1}$ | diffusivity in NMC111 | [6], [a] |
| $k_0$ | $10^{-10}\ \text{mol}^{\alpha-1}\text{m}^{4-3\alpha}\text{s}^{-1}$ | reaction rate constant | [1] |
| $\alpha$ | 0.5 | charge transfer coefficient | assumption |
| $\nu_\text{tdf}$ | Eq. (29) | thermodynamic factor | [5] |
| $t_+$ | 0.23 | transference number of the lithium ion | [1], [5] |
| $c_\text{s,max}$ | Eq. (30) | maximum concentration in NMC111 | [1] |
| $c_\text{s,0}$ | Eq. (31) | initial concentration in NMC111 | [1] |
| $\bar{c}_\text{e,0}$ | $1000\ \text{mol}\,\text{m}^{-3}$ | initial concentration in the electrolyte | |
| $\rho_\text{s}$ | $4770\ \text{kg}\,\text{m}^{-3}$ | density of NMC111 | [7] |
| $\rho_\text{bsg}$ | $2.23\ \text{g}\,\text{cm}^{-3}$ | density of borosilicate glass | [8] |
| $w^\text{sep}$ | $53\ \text{g}\,\text{cm}^{-2}$ | area density of the separator | [2] |
| $Q_\text{spec}$ | $278\ \text{mA}\,\text{h}\,\text{g}^{-1}$ | specific nominal charge capacity | [9] |
| $Q_\text{rev}$ | $164\ \text{mA}\,\text{h}\,\text{g}^{-1}$ | specific reversibly accessible charge capacity | |
| $T$ | $298\ \text{K}$ | temperature | |

[a] Fitting parameter.

## 2.2 Effective transport properties

The M-factor is defined as the ratio of effective over intrinsic transport property, e.g. diffusivity or conductivity. Formulas allowing the prediction of the M-factor based on purely morphological descriptors are available[10]. One of these formulas also applies to the considered nanostructured NMC111 secondary particles [11]. However, these formulas require 3D morphological information, which is experimentally not accessible. This includes descriptors quantifying the minimum length of transport paths and the degree of bottleneck effects. In order to predict the M-factors of nanostructured NMC111 particles, for which no image data is available, we fit a Bruggemann-type formula to the data presented by Neumann et al. [11]. This means that we assume the M-factor for ionic or electronic transport to depend on the volume fraction $\varepsilon$ of pores or active material as follows: $M = \varepsilon^\alpha$. The parameter $\alpha$ is determined by the method of least squares. We obtain $\alpha = 1.801$ and $\alpha = 2.140$ for ionic and electronic transport, respectively. The goodness-of-fit is visualized in 2.

Table 3: Effective transport properties of the reference cell.

| Parameter | Value | Description | Source |
|---|---|---|---|
| $\kappa_{\text{eff}}$ | $M\kappa$ | effective transport property | |
| $M_{\text{ion}}^{\text{sep}}$ | $(\varepsilon_{\text{e}}^{\text{sep}})^{1.42}$ | M-factor for ionic transport in the separator | [12] |
| $M_{\text{ion}}^{\text{pos}}$ | $\varepsilon_{\text{e}}^{\text{pos}}(\varepsilon_{\text{e}}^{\text{pos}} + \varepsilon_{\text{f}}^{\text{pos}})^{0.342}$ | M-factor for ionic transport in the positive electrode | [13, 1] |
| $M_{\text{eon}}^{\text{pos}}$ | $\varepsilon_{\text{f}}^{\text{pos}}(\varepsilon_{\text{e}}^{\text{pos}} + \varepsilon_{\text{f}}^{\text{pos}})^{0.342}$ | M-factor for electronic transport in the positive electrode | [13, 1] |
| $M_{\text{ion}}^{\text{II}}$ | $\left(\varepsilon_{\text{e}}^{(\text{II})}\right)^{1.801}$ | M-factor for ionic transport in the secondary particles | [11] |
| $M_{\text{eon}}^{\text{II}}$ | $\left(\varepsilon_{\text{s}}^{(\text{II})}\right)^{2.140}$ | M-factor for electronic transport in the secondary particles | [11] |

# References

[1] Oleg Birkholz and Marc Kamlah. Electrochemical Modeling of Hierarchically Structured Lithium-Ion Battery Electrodes. Energy Technol., 9(6):2000910, 2021.

[2] FORTUNE Media IP Limited VWR International, LLC. Glass microfibre filters without binder, grade GF/C, Whatman, 2021.

[3] Sarah G. Stewart, Venkat Srinivasan, and John Newman. Modeling the Performance of Lithium-Ion Batteries and Capacitors during Hybrid-Electric-Vehicle Operation. J. Electrochem. Soc., 155(9):A664, July 2008.

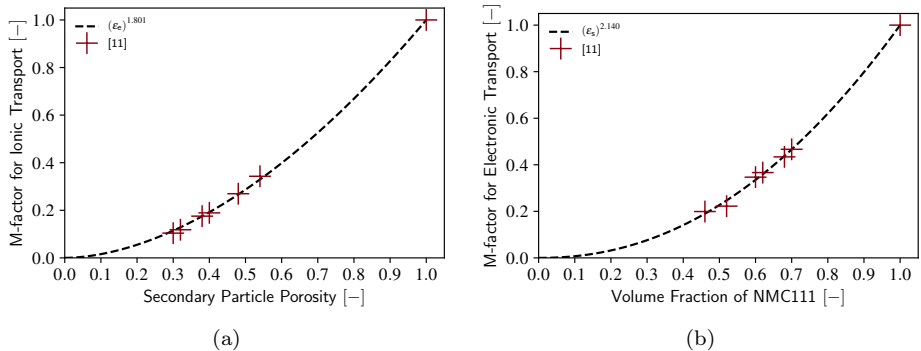

Figure 2: Fit of M-factor for (a) ionic and (b) electronic transport in the secondary particle.

[4] Julian Zahnow, Tim Bernges, Amalia Wagner, Nicole Bohn, Joachim R. Binder, Wolfgang G. Zeier, Matthias T. Elm, and Jürgen Janek. Impedance Analysis of NCM Cathode Materials: Electronic and Ionic Partial Conductivities and the Influence of Microstructure. ACS Appl. Energy Mater., 4(2):1335–1345, February 2021.

[5] Johannes Landesfeind and Hubert A. Gasteiger. Temperature and Concentration Dependence of the Ionic Transport Properties of Lithium-Ion Battery Electrolytes. J. Electrochem. Soc., 166(14):A3079, September 2019.

[6] Shao-Ling Wu, Wei Zhang, Xiangyun Song, Alpesh K. Shukla, Gao Liu, Vincent Battaglia, and Venkat Srinivasan. High Rate Capability of Li(Ni1/3Mn1/3Co1/3)O2 Electrode for Li-Ion Batteries. J. Electrochem. Soc., 159(4):A438–A444, January 2012.

[7] S.-C. Yin, Y.-H. Rho, I. Swainson, and L. F. Nazar. X-ray/Neutron Diffraction and Electrochemical Studies of Lithium De/Re-Intercalation in Li1-xCo1/3Ni1/3Mn1/3O2 (x = 0 → 1). Chem. Mater., 18(7):1901–1910, April 2006.

[8] A. R. Boccaccini, D. R. Acevedo, G. Brusatin, and P. Colombo. Borosilicate glass matrix composites containing multi-wall carbon nanotubes. J. Eur. Ceram., 25(9):1515–1523, June 2005.

[9] Oleg Birkholz. Modeling transport properties and electrochemical performance of hierarchically structured lithium-ion battery cathodes using resistor networks and mathematical half-cell models. KIT Scientific Publishing, October 2022.

[10] M. Neumann, O. Stenzel, F. Willot, L. Holzer, and V. Schmidt. Quantifying the influence of microstructure on effective conductivity and permeability: Virtual materials testing. Int. J. Solids Struct., 184:211–220, 2020.

[11] Matthias Neumann, Sven E. Wetterauer, Markus Osenberg, André Hilger, Phillip Gräfensteiner, Amalia Wagner, Nicole Bohn, Joachim R. Binder, Ingo Manke, Thomas Carraro, and Volker Schmidt. A Data-Driven Modeling Approach to Quantify Morphology Effects on Transport Properties in Nanostructured NMC Particles. preprint, SSRN, 2023.

[12] Roland Balbierer. Charakterisierung von Lithium-Ionen-Batterien und deren Komponenten mittels NMR-Methoden, 2021.

[13] Oleg Birkholz, Matthias Neumann, Volker Schmidt, and Marc Kamlah. Statistical investigation of structural and transport properties of densely-packed assemblies of overlapping spheres using the resistor network method. Powder Technol., 378:659–666, January 2021.



\end{document}